\documentclass[sigplan,authorversion]{acmart}

\clearpage{}%
\usepackage[utf8]{inputenc}
\usepackage{textgreek}
\usepackage{fixltx2e} %
\usepackage{ifpdf}
\usepackage{graphicx}

\newif\ifhtml
\ifpdf
\htmlfalse
\else
\htmltrue
\fi

\usepackage{scripts/collab}  %

\newcommand{\ie}{i.e.\xspace}
\newcommand{\eg}{e.g.\xspace}

\usepackage{letltxmacro}
\LetLtxMacro{\OrgCitep}{\citep}
\renewcommand{\citep}[1]{\,\OrgCitep{#1}}
\renewcommand{\cite}[1]{\PackageError{misc-macros}{Please do not use cite macro. Was used as \ cite{#1}}{Please use \ citep{#1} or \ citet{#1} instead.}}

\newcommand{\code}[1]{\texttt{#1}}
\usepackage{listings}
\usepackage{ifthen}
\newcommand{\citeurl}[5]{%
#1\footnote{\emph{#2}%
          \ifthenelse{\equal{#3}{}}
                     {}%
                     {, #3}%
          \ifthenelse{\equal{#4}{}}
                     {}%
                     {, access date: #4}%
          , \url{#5}}}

\newcommand{\inlinequote}[1]{\emph{``#1''}}

\newcommand{\circled}[1]{\textcircled{\scriptsize\textsf{#1}}}

\makeatletter
\@ifpackageloaded{tex4ht}
  {\newcommand\knitrForHt{}}%
  {\usepackage{tikz}
\tikzstyle{every picture}+=[font=\sffamily]}
\makeatother

\definecolor{fgcolor}{rgb}{0.345, 0.345, 0.345}
\definecolor{messagecolor}{rgb}{0, 0, 0}
\definecolor{errorcolor}{rgb}{1, 0, 0}

\newenvironment{knitrout}{}{} %

\usepackage{framed}
\makeatletter
 {\par\unskip\endMakeFramed%
 \at@end@of@kframe}
\makeatother
\usepackage{alltt}
 
\usepackage[nameinlink]{cleveref}
\Crefformat{lstlisting}{#2Lst.\,#1#3}
\crefformat{lstlisting}{#2lst.\,#1#3}

\clearpage{}%
\clearpage{}%

\def\SavinaSnapshots{%
\begin{knitrout}
\definecolor{shadecolor}{rgb}{0.969, 0.969, 0.969}

\end{knitrout}
}%

\newcommand{\SavinaMinOverheadP}{
  0.04\%\xspace}

\newcommand{\SavinaMaxOverheadP}{
  400.63\%\xspace}

\newcommand{\SavinaMaxOverhead}{
  5x\xspace}

\newcommand{\SavinaAvgOverheadP}{
  49.35\%\xspace}

\newcommand{\SavinaRRAvgOverheadP}{
  54.88\%\xspace}

\newcommand{\SavinaFJTGCOverheadShareP}{
  39.82\%\xspace}

\newcommand{\SavinaFJTGCOH}{
  19.22ms\xspace}

\newcommand{\SavinaFJTtime}{
  40.36ms\xspace}
\clearpage{}%
\clearpage{}%

\def\AwfyBaseline{%
\begin{knitrout}
\definecolor{shadecolor}{rgb}{0.969, 0.969, 0.969}
\begin{tikzpicture}[x=1pt,y=1pt]
\definecolor{fillColor}{RGB}{255,255,255}
\path[use as bounding box,fill=fillColor,fill opacity=0.00] (0,0) rectangle (101.18,158.99);
\begin{scope}
\path[clip] ( 27.40, 30.99) rectangle (101.18,158.99);
\definecolor{drawColor}{gray}{0.80}

\path[draw=drawColor,line width= 0.6pt,dash pattern=on 4pt off 4pt ,line join=round] ( 27.40, 42.85) -- (101.18, 42.85);

\path[draw=drawColor,line width= 0.6pt,dash pattern=on 4pt off 4pt ,line join=round] ( 27.40, 77.33) -- (101.18, 77.33);

\path[draw=drawColor,line width= 0.6pt,dash pattern=on 4pt off 4pt ,line join=round] ( 27.40,111.81) -- (101.18,111.81);
\definecolor{drawColor}{gray}{0.20}

\path[draw=drawColor,line width= 0.6pt,line join=round] ( 41.24, 42.85) -- ( 41.24, 42.85);

\path[draw=drawColor,line width= 0.6pt,line join=round] ( 41.24, 42.85) -- ( 41.24, 42.85);
\definecolor{fillColor}{RGB}{255,255,255}

\path[draw=drawColor,line width= 0.6pt,line join=round,line cap=round,fill=fillColor] ( 32.59, 42.85) --
	( 32.59, 42.85) --
	( 49.88, 42.85) --
	( 49.88, 42.85) --
	( 32.59, 42.85) --
	cycle;

\path[draw=drawColor,line width= 1.1pt,line join=round] ( 32.59, 42.85) -- ( 49.88, 42.85);

\path[draw=drawColor,line width= 0.6pt,line join=round] ( 64.29, 85.99) -- ( 64.29,106.29);

\path[draw=drawColor,line width= 0.6pt,line join=round] ( 64.29, 58.67) -- ( 64.29, 36.81);

\path[draw=drawColor,line width= 0.6pt,line join=round,line cap=round,fill=fillColor] ( 55.65, 85.99) --
	( 55.65, 58.67) --
	( 72.94, 58.67) --
	( 72.94, 85.99) --
	( 55.65, 85.99) --
	cycle;

\path[draw=drawColor,line width= 1.1pt,line join=round] ( 55.65, 73.46) -- ( 72.94, 73.46);

\path[draw=drawColor,line width= 0.6pt,line join=round] ( 87.35,104.54) -- ( 87.35,153.18);

\path[draw=drawColor,line width= 0.6pt,line join=round] ( 87.35, 69.14) -- ( 87.35, 41.58);

\path[draw=drawColor,line width= 0.6pt,line join=round,line cap=round,fill=fillColor] ( 78.70,104.54) --
	( 78.70, 69.14) --
	( 95.99, 69.14) --
	( 95.99,104.54) --
	( 78.70,104.54) --
	cycle;

\path[draw=drawColor,line width= 1.1pt,line join=round] ( 78.70, 76.76) -- ( 95.99, 76.76);
\end{scope}
\begin{scope}
\path[clip] (  0.00,  0.00) rectangle (101.18,158.99);
\definecolor{drawColor}{RGB}{190,190,190}

\path[draw=drawColor,line width= 0.6pt,line join=round] ( 27.40, 30.99) --
	( 27.40,158.99);
\end{scope}
\begin{scope}
\path[clip] (  0.00,  0.00) rectangle (101.18,158.99);
\definecolor{drawColor}{gray}{0.30}

\node[text=drawColor,anchor=base east,inner sep=0pt, outer sep=0pt, scale=  0.80] at ( 22.45, 40.09) {1};

\node[text=drawColor,anchor=base east,inner sep=0pt, outer sep=0pt, scale=  0.80] at ( 22.45, 74.58) {2};

\node[text=drawColor,anchor=base east,inner sep=0pt, outer sep=0pt, scale=  0.80] at ( 22.45,109.06) {3};

\node[text=drawColor,anchor=base east,inner sep=0pt, outer sep=0pt, scale=  0.80] at ( 22.45,143.54) {4};
\end{scope}
\begin{scope}
\path[clip] (  0.00,  0.00) rectangle (101.18,158.99);
\definecolor{drawColor}{gray}{0.20}

\path[draw=drawColor,line width= 0.6pt,line join=round] ( 24.65, 42.85) --
	( 27.40, 42.85);

\path[draw=drawColor,line width= 0.6pt,line join=round] ( 24.65, 77.33) --
	( 27.40, 77.33);

\path[draw=drawColor,line width= 0.6pt,line join=round] ( 24.65,111.81) --
	( 27.40,111.81);

\path[draw=drawColor,line width= 0.6pt,line join=round] ( 24.65,146.30) --
	( 27.40,146.30);
\end{scope}
\begin{scope}
\path[clip] (  0.00,  0.00) rectangle (101.18,158.99);
\definecolor{drawColor}{RGB}{190,190,190}

\path[draw=drawColor,line width= 0.6pt,line join=round] ( 27.40, 30.99) --
	(101.18, 30.99);
\end{scope}
\begin{scope}
\path[clip] (  0.00,  0.00) rectangle (101.18,158.99);
\definecolor{drawColor}{gray}{0.20}

\path[draw=drawColor,line width= 0.6pt,line join=round] ( 41.24, 28.24) --
	( 41.24, 30.99);

\path[draw=drawColor,line width= 0.6pt,line join=round] ( 64.29, 28.24) --
	( 64.29, 30.99);

\path[draw=drawColor,line width= 0.6pt,line join=round] ( 87.35, 28.24) --
	( 87.35, 30.99);
\end{scope}
\begin{scope}
\path[clip] (  0.00,  0.00) rectangle (101.18,158.99);
\definecolor{drawColor}{gray}{0.30}

\node[text=drawColor,rotate= 90.00,anchor=base east,inner sep=0pt, outer sep=0pt, scale=  0.80] at ( 43.99, 26.04) {Java};

\node[text=drawColor,rotate= 90.00,anchor=base east,inner sep=0pt, outer sep=0pt, scale=  0.80] at ( 67.05, 26.04) {Node.js};

\node[text=drawColor,rotate= 90.00,anchor=base east,inner sep=0pt, outer sep=0pt, scale=  0.80] at ( 90.10, 26.04) {SOMns};
\end{scope}
\begin{scope}
\path[clip] (  0.00,  0.00) rectangle (101.18,158.99);
\definecolor{drawColor}{RGB}{0,0,0}

\node[text=drawColor,rotate= 90.00,anchor=base,inner sep=0pt, outer sep=0pt, scale=  0.80] at (  5.51, 94.99) {Runtime Factor nnormalized to Java};

\node[text=drawColor,rotate= 90.00,anchor=base,inner sep=0pt, outer sep=0pt, scale=  0.80] at ( 14.15, 94.99) {(lower is better)};
\end{scope}
\end{tikzpicture}
 
\end{knitrout}
}%

\def\SavinaBaseline{%
\begin{knitrout}
\definecolor{shadecolor}{rgb}{0.969, 0.969, 0.969}

\end{knitrout}
}%

\def\AcmeBox{%
\begin{knitrout}
\definecolor{shadecolor}{rgb}{0.969, 0.969, 0.969}
\begin{tikzpicture}[x=1pt,y=1pt]
\definecolor{fillColor}{RGB}{255,255,255}
\path[use as bounding box,fill=fillColor,fill opacity=0.00] (0,0) rectangle (166.22,144.54);
\begin{scope}
\path[clip] ( 77.04, 47.51) rectangle (166.22,144.54);
\definecolor{drawColor}{gray}{0.20}

\path[draw=drawColor,line width= 0.6pt,line join=round] (114.65, 54.61) -- (125.84, 54.61);

\path[draw=drawColor,line width= 0.6pt,line join=round] ( 92.28, 54.61) -- ( 81.09, 54.61);
\definecolor{fillColor}{RGB}{255,255,255}

\path[draw=drawColor,line width= 0.6pt,line join=round,line cap=round,fill=fillColor] (114.65, 50.18) --
	( 92.28, 50.18) --
	( 92.28, 59.05) --
	(114.65, 59.05) --
	(114.65, 50.18) --
	cycle;

\path[draw=drawColor,line width= 1.1pt,line join=round] (103.47, 50.18) -- (103.47, 59.05);

\path[draw=drawColor,line width= 0.6pt,line join=round] (123.78, 66.44) -- (138.01, 66.44);

\path[draw=drawColor,line width= 0.6pt,line join=round] ( 95.32, 66.44) -- ( 81.09, 66.44);

\path[draw=drawColor,line width= 0.6pt,line join=round,line cap=round,fill=fillColor] (123.78, 62.01) --
	( 95.32, 62.01) --
	( 95.32, 70.88) --
	(123.78, 70.88) --
	(123.78, 62.01) --
	cycle;

\path[draw=drawColor,line width= 1.1pt,line join=round] (109.55, 62.01) -- (109.55, 70.88);

\path[draw=drawColor,line width= 0.6pt,line join=round] (125.56, 78.28) -- (140.38, 78.28);

\path[draw=drawColor,line width= 0.6pt,line join=round] ( 95.91, 78.28) -- ( 81.09, 78.28);

\path[draw=drawColor,line width= 0.6pt,line join=round,line cap=round,fill=fillColor] (125.56, 73.84) --
	( 95.91, 73.84) --
	( 95.91, 82.71) --
	(125.56, 82.71) --
	(125.56, 73.84) --
	cycle;

\path[draw=drawColor,line width= 1.1pt,line join=round] (110.73, 73.84) -- (110.73, 82.71);

\path[draw=drawColor,line width= 0.6pt,line join=round] (108.94, 90.11) -- (118.23, 90.11);

\path[draw=drawColor,line width= 0.6pt,line join=round] ( 90.38, 90.11) -- ( 81.09, 90.11);

\path[draw=drawColor,line width= 0.6pt,line join=round,line cap=round,fill=fillColor] (108.94, 85.67) --
	( 90.38, 85.67) --
	( 90.38, 94.55) --
	(108.94, 94.55) --
	(108.94, 85.67) --
	cycle;

\path[draw=drawColor,line width= 1.1pt,line join=round] ( 99.66, 85.67) -- ( 99.66, 94.55);

\path[draw=drawColor,line width= 0.6pt,line join=round] (136.67,101.94) -- (155.19,101.94);

\path[draw=drawColor,line width= 0.6pt,line join=round] ( 99.62,101.94) -- ( 81.09,101.94);

\path[draw=drawColor,line width= 0.6pt,line join=round,line cap=round,fill=fillColor] (136.67, 97.51) --
	( 99.62, 97.51) --
	( 99.62,106.38) --
	(136.67,106.38) --
	(136.67, 97.51) --
	cycle;

\path[draw=drawColor,line width= 1.1pt,line join=round] (118.14, 97.51) -- (118.14,106.38);

\path[draw=drawColor,line width= 0.6pt,line join=round] (129.71,113.78) -- (145.92,113.78);

\path[draw=drawColor,line width= 0.6pt,line join=round] ( 97.30,113.78) -- ( 81.09,113.78);

\path[draw=drawColor,line width= 0.6pt,line join=round,line cap=round,fill=fillColor] (129.71,109.34) --
	( 97.30,109.34) --
	( 97.30,118.21) --
	(129.71,118.21) --
	(129.71,109.34) --
	cycle;

\path[draw=drawColor,line width= 1.1pt,line join=round] (113.51,109.34) -- (113.51,118.21);

\path[draw=drawColor,line width= 0.6pt,line join=round] (112.16,125.61) -- (122.51,125.61);

\path[draw=drawColor,line width= 0.6pt,line join=round] ( 91.45,125.61) -- ( 81.09,125.61);

\path[draw=drawColor,line width= 0.6pt,line join=round,line cap=round,fill=fillColor] (112.16,121.17) --
	( 91.45,121.17) --
	( 91.45,130.05) --
	(112.16,130.05) --
	(112.16,121.17) --
	cycle;

\path[draw=drawColor,line width= 1.1pt,line join=round] (101.80,121.17) -- (101.80,130.05);

\path[draw=drawColor,line width= 0.6pt,line join=round] (120.62,137.44) -- (133.80,137.44);

\path[draw=drawColor,line width= 0.6pt,line join=round] ( 94.27,137.44) -- ( 81.09,137.44);

\path[draw=drawColor,line width= 0.6pt,line join=round,line cap=round,fill=fillColor] (120.62,133.00) --
	( 94.27,133.00) --
	( 94.27,141.88) --
	(120.62,141.88) --
	(120.62,133.00) --
	cycle;

\path[draw=drawColor,line width= 1.1pt,line join=round] (107.45,133.00) -- (107.45,141.88);
\definecolor{drawColor}{gray}{0.80}

\path[draw=drawColor,line width= 0.6pt,dash pattern=on 4pt off 4pt ,line join=round] ( 81.09, 47.51) -- ( 81.09,144.54);
\end{scope}
\begin{scope}
\path[clip] (  0.00,  0.00) rectangle (166.22,144.54);
\definecolor{drawColor}{RGB}{190,190,190}

\path[draw=drawColor,line width= 0.6pt,line join=round] ( 77.04, 47.51) --
	( 77.04,144.54);
\end{scope}
\begin{scope}
\path[clip] (  0.00,  0.00) rectangle (166.22,144.54);
\definecolor{drawColor}{gray}{0.30}

\node[text=drawColor,anchor=base east,inner sep=0pt, outer sep=0pt, scale=  0.80] at ( 72.09, 51.86) {BookFlight};

\node[text=drawColor,anchor=base east,inner sep=0pt, outer sep=0pt, scale=  0.80] at ( 72.09, 63.69) {Cancel Booking};

\node[text=drawColor,anchor=base east,inner sep=0pt, outer sep=0pt, scale=  0.80] at ( 72.09, 75.52) {List Bookings};

\node[text=drawColor,anchor=base east,inner sep=0pt, outer sep=0pt, scale=  0.80] at ( 72.09, 87.36) {Login};

\node[text=drawColor,anchor=base east,inner sep=0pt, outer sep=0pt, scale=  0.80] at ( 72.09, 99.19) {Query Flight};

\node[text=drawColor,anchor=base east,inner sep=0pt, outer sep=0pt, scale=  0.80] at ( 72.09,111.02) {Update Customer};

\node[text=drawColor,anchor=base east,inner sep=0pt, outer sep=0pt, scale=  0.80] at ( 72.09,122.85) {View Profile};

\node[text=drawColor,anchor=base east,inner sep=0pt, outer sep=0pt, scale=  0.80] at ( 72.09,134.69) {Logout};
\end{scope}
\begin{scope}
\path[clip] (  0.00,  0.00) rectangle (166.22,144.54);
\definecolor{drawColor}{gray}{0.20}

\path[draw=drawColor,line width= 0.6pt,line join=round] ( 74.29, 54.61) --
	( 77.04, 54.61);

\path[draw=drawColor,line width= 0.6pt,line join=round] ( 74.29, 66.44) --
	( 77.04, 66.44);

\path[draw=drawColor,line width= 0.6pt,line join=round] ( 74.29, 78.28) --
	( 77.04, 78.28);

\path[draw=drawColor,line width= 0.6pt,line join=round] ( 74.29, 90.11) --
	( 77.04, 90.11);

\path[draw=drawColor,line width= 0.6pt,line join=round] ( 74.29,101.94) --
	( 77.04,101.94);

\path[draw=drawColor,line width= 0.6pt,line join=round] ( 74.29,113.78) --
	( 77.04,113.78);

\path[draw=drawColor,line width= 0.6pt,line join=round] ( 74.29,125.61) --
	( 77.04,125.61);

\path[draw=drawColor,line width= 0.6pt,line join=round] ( 74.29,137.44) --
	( 77.04,137.44);
\end{scope}
\begin{scope}
\path[clip] (  0.00,  0.00) rectangle (166.22,144.54);
\definecolor{drawColor}{RGB}{190,190,190}

\path[draw=drawColor,line width= 0.6pt,line join=round] ( 77.04, 47.51) --
	(166.22, 47.51);
\end{scope}
\begin{scope}
\path[clip] (  0.00,  0.00) rectangle (166.22,144.54);
\definecolor{drawColor}{gray}{0.20}

\path[draw=drawColor,line width= 0.6pt,line join=round] ( 81.09, 44.76) --
	( 81.09, 47.51);

\path[draw=drawColor,line width= 0.6pt,line join=round] ( 97.31, 44.76) --
	( 97.31, 47.51);

\path[draw=drawColor,line width= 0.6pt,line join=round] (113.52, 44.76) --
	(113.52, 47.51);

\path[draw=drawColor,line width= 0.6pt,line join=round] (129.74, 44.76) --
	(129.74, 47.51);

\path[draw=drawColor,line width= 0.6pt,line join=round] (145.95, 44.76) --
	(145.95, 47.51);

\path[draw=drawColor,line width= 0.6pt,line join=round] (162.17, 44.76) --
	(162.17, 47.51);
\end{scope}
\begin{scope}
\path[clip] (  0.00,  0.00) rectangle (166.22,144.54);
\definecolor{drawColor}{gray}{0.30}

\node[text=drawColor,rotate= 90.00,anchor=base east,inner sep=0pt, outer sep=0pt, scale=  0.80] at ( 83.85, 42.56) {1.000};

\node[text=drawColor,rotate= 90.00,anchor=base east,inner sep=0pt, outer sep=0pt, scale=  0.80] at (100.06, 42.56) {1.005};

\node[text=drawColor,rotate= 90.00,anchor=base east,inner sep=0pt, outer sep=0pt, scale=  0.80] at (116.28, 42.56) {1.010};

\node[text=drawColor,rotate= 90.00,anchor=base east,inner sep=0pt, outer sep=0pt, scale=  0.80] at (132.49, 42.56) {1.015};

\node[text=drawColor,rotate= 90.00,anchor=base east,inner sep=0pt, outer sep=0pt, scale=  0.80] at (148.71, 42.56) {1.020};

\node[text=drawColor,rotate= 90.00,anchor=base east,inner sep=0pt, outer sep=0pt, scale=  0.80] at (164.92, 42.56) {1.025};
\end{scope}
\begin{scope}
\path[clip] (  0.00,  0.00) rectangle (166.22,144.54);
\definecolor{drawColor}{RGB}{0,0,0}

\node[text=drawColor,anchor=base,inner sep=0pt, outer sep=0pt, scale=  0.80] at (121.63, 18.06) {Latency Factor};

\node[text=drawColor,anchor=base,inner sep=0pt, outer sep=0pt, scale=  0.80] at (121.63,  9.42) {normalized to baseline};

\node[text=drawColor,anchor=base,inner sep=0pt, outer sep=0pt, scale=  0.80] at (121.63,  0.78) {(lower is better)};
\end{scope}
\end{tikzpicture}
 
\end{knitrout}
}%

\newcommand{\AcmeLatencyTracedMaxX}{
  2.28\%\xspace}

\newcommand{\AcmeLatencySnapCntGFifty}{
  12821\xspace}

\newcommand{\AcmeLatencySnapPSFifty}{
  99.93\%\xspace}

\newcommand{\AcmeLatencySnapCntGHundred}{
  1359\xspace}

\newcommand{\AcmeLatencySnapCntGTwohundred}{
  385\xspace}

\newcommand{\AcmeLatencySnapMax}{
  625ms\xspace}

\newcommand{\AcmeLatencyOriginalCntGFifty}{
  16611\xspace}

\newcommand{\AcmeLatencyOriginalCntGHundred}{
  1289\xspace}

\newcommand{\AcmeLatencyOriginalCntGTwohundred}{
  250\xspace}

\newcommand{\AcmeLatencyIncreaseHundredFHundred}{
  7.53\%\xspace}
              
\newcommand{\AcmeLatencyIncreaseHundred}{
  5.43\%\xspace}

\newcommand{\AcmeLatencySHundredShare}{
  0.007\%\xspace}
\clearpage{}%
\clearpage{}%

\def\ACDCBox{%
\begin{knitrout}
\definecolor{shadecolor}{rgb}{0.969, 0.969, 0.969}
\begin{tikzpicture}[x=1pt,y=1pt]
\definecolor{fillColor}{RGB}{255,255,255}
\path[use as bounding box,fill=fillColor,fill opacity=0.00] (0,0) rectangle (108.41,158.99);
\begin{scope}
\path[clip] ( 35.10, 30.54) rectangle (108.41,158.99);
\definecolor{drawColor}{gray}{0.20}

\path[draw=drawColor,line width= 0.6pt,line join=round] ( 55.09,147.31) -- ( 55.09,153.16);

\path[draw=drawColor,line width= 0.6pt,line join=round] ( 55.09,139.78) -- ( 55.09,130.52);
\definecolor{fillColor}{RGB}{255,255,255}

\path[draw=drawColor,line width= 0.6pt,line join=round,line cap=round,fill=fillColor] ( 42.59,147.31) --
	( 42.59,139.78) --
	( 67.59,139.78) --
	( 67.59,147.31) --
	( 42.59,147.31) --
	cycle;

\path[draw=drawColor,line width= 1.1pt,line join=round] ( 42.59,144.96) -- ( 67.59,144.96);
\definecolor{fillColor}{gray}{0.20}

\path[draw=drawColor,line width= 0.4pt,line join=round,line cap=round,fill=fillColor] ( 88.41, 36.38) circle (  0.89);

\path[draw=drawColor,line width= 0.4pt,line join=round,line cap=round,fill=fillColor] ( 88.41, 36.38) circle (  0.89);

\path[draw=drawColor,line width= 0.4pt,line join=round,line cap=round,fill=fillColor] ( 88.41, 36.38) circle (  0.89);

\path[draw=drawColor,line width= 0.4pt,line join=round,line cap=round,fill=fillColor] ( 88.41, 36.38) circle (  0.89);

\path[draw=drawColor,line width= 0.4pt,line join=round,line cap=round,fill=fillColor] ( 88.41, 36.38) circle (  0.89);

\path[draw=drawColor,line width= 0.6pt,line join=round] ( 88.41, 36.38) -- ( 88.41, 36.38);

\path[draw=drawColor,line width= 0.6pt,line join=round] ( 88.41, 36.38) -- ( 88.41, 36.38);
\definecolor{fillColor}{RGB}{255,255,255}

\path[draw=drawColor,line width= 0.6pt,line join=round,line cap=round,fill=fillColor] ( 75.92, 36.38) --
	( 75.92, 36.38) --
	(100.91, 36.38) --
	(100.91, 36.38) --
	( 75.92, 36.38) --
	cycle;

\path[draw=drawColor,line width= 1.1pt,line join=round] ( 75.92, 36.38) -- (100.91, 36.38);
\definecolor{drawColor}{gray}{0.80}

\path[draw=drawColor,line width= 0.6pt,dash pattern=on 4pt off 4pt ,line join=round] ( 35.10, 36.38) -- (108.41, 36.38);

\path[draw=drawColor,line width= 0.6pt,dash pattern=on 4pt off 4pt ,line join=round] ( 35.10,146.91) -- (108.41,146.91);
\end{scope}
\begin{scope}
\path[clip] (  0.00,  0.00) rectangle (108.41,158.99);
\definecolor{drawColor}{RGB}{190,190,190}

\path[draw=drawColor,line width= 0.6pt,line join=round] ( 35.10, 30.54) --
	( 35.10,158.99);
\end{scope}
\begin{scope}
\path[clip] (  0.00,  0.00) rectangle (108.41,158.99);
\definecolor{drawColor}{gray}{0.30}

\node[text=drawColor,anchor=base east,inner sep=0pt, outer sep=0pt, scale=  0.80] at ( 30.15, 33.63) {1};

\node[text=drawColor,anchor=base east,inner sep=0pt, outer sep=0pt, scale=  0.80] at ( 30.15, 67.93) {10};

\node[text=drawColor,anchor=base east,inner sep=0pt, outer sep=0pt, scale=  0.80] at ( 30.15,106.04) {20};

\node[text=drawColor,anchor=base east,inner sep=0pt, outer sep=0pt, scale=  0.80] at ( 30.15,144.16) {30};
\end{scope}
\begin{scope}
\path[clip] (  0.00,  0.00) rectangle (108.41,158.99);
\definecolor{drawColor}{gray}{0.20}

\path[draw=drawColor,line width= 0.6pt,line join=round] ( 32.35, 36.38) --
	( 35.10, 36.38);

\path[draw=drawColor,line width= 0.6pt,line join=round] ( 32.35, 70.68) --
	( 35.10, 70.68);

\path[draw=drawColor,line width= 0.6pt,line join=round] ( 32.35,108.80) --
	( 35.10,108.80);

\path[draw=drawColor,line width= 0.6pt,line join=round] ( 32.35,146.91) --
	( 35.10,146.91);
\end{scope}
\begin{scope}
\path[clip] (  0.00,  0.00) rectangle (108.41,158.99);
\definecolor{drawColor}{RGB}{190,190,190}

\path[draw=drawColor,line width= 0.6pt,line join=round] ( 35.10, 30.54) --
	(108.41, 30.54);
\end{scope}
\begin{scope}
\path[clip] (  0.00,  0.00) rectangle (108.41,158.99);
\definecolor{drawColor}{gray}{0.20}

\path[draw=drawColor,line width= 0.6pt,line join=round] ( 55.09, 27.79) --
	( 55.09, 30.54);

\path[draw=drawColor,line width= 0.6pt,line join=round] ( 88.41, 27.79) --
	( 88.41, 30.54);
\end{scope}
\begin{scope}
\path[clip] (  0.00,  0.00) rectangle (108.41,158.99);
\definecolor{drawColor}{gray}{0.30}

\node[text=drawColor,rotate= 90.00,anchor=base east,inner sep=0pt, outer sep=0pt, scale=  0.80] at ( 57.85, 25.59) {Fuel};

\node[text=drawColor,rotate= 90.00,anchor=base east,inner sep=0pt, outer sep=0pt, scale=  0.80] at ( 91.17, 25.59) {SOMns};
\end{scope}
\begin{scope}
\path[clip] (  0.00,  0.00) rectangle (108.41,158.99);
\definecolor{drawColor}{RGB}{0,0,0}

\node[text=drawColor,rotate= 90.00,anchor=base,inner sep=0pt, outer sep=0pt, scale=  0.80] at (  5.51, 94.77) {Runtime Factor nnormalized to Java};

\node[text=drawColor,rotate= 90.00,anchor=base,inner sep=0pt, outer sep=0pt, scale=  0.80] at ( 14.15, 94.77) {(lower is better)};
\end{scope}
\end{tikzpicture}
 
\end{knitrout}
}%

\newcommand{\ACDCPharoMaxX}{
  31.64\xspace}

\newcommand{\ACDCPharoGMeanX}{
  29.1\xspace}
\clearpage{}%

\usepackage{subcaption}
\usepackage{listings}

\def\SOMns{SOM{\sc ns}\xspace}
\def\AcmeAir{Acme\,Air\xspace}
\def\Kompos{Kómpos\xspace}

\def\BibTeX{{\rm B\kern-.05em{\sc i\kern-.025em b}\kern-.08emT\kern-.1667em\lower.7ex\hbox{E}\kern-.125emX}}

\setcopyright{acmlicensed}
\copyrightyear{2019} 
\acmYear{2019} 
\acmConference[MPLR '19]{Proceedings of the 16th ACM SIGPLAN International Conference on Managed Programming Languages and Runtimes}{October 21--22, 2019}{Athens, Greece}
\acmBooktitle{Proceedings of the 16th ACM SIGPLAN International Conference on Managed Programming Languages and Runtimes (MPLR '19), October 21--22, 2019, Athens, Greece}
\acmPrice{15.00}
\acmDOI{10.1145/3357390.3361019}
\acmISBN{978-1-4503-6977-0/19/10}

\collabAuthor{sm}{red}{Stefan}
\collabAuthor{egb}{blue}{Elisa}
\collabAuthor{ctl}{teal}{Carmen}
\collabAuthor{hm}{violet}{Hanspeter}
\collabAuthor{dau}{orange}{Dominik}

\begin{document}

\title[Asynchronous Snapshots of Actor Systems]{Asynchronous Snapshots of Actor Systems for Latency-Sensitive Applications}
\author{Dominik Aumayr}
\affiliation{%
  \institution{Johannes Kepler University}
  \city{Linz}
  \country{Austria}
}
\email{dominik.aumayr@jku.at}

\author{Stefan Marr}
\affiliation{%
  \institution{University of Kent}
  \city{Canterbury}
  \country{United Kingdom}
}
\email{s.marr@kent.ac.uk}

\author{Elisa Gonzalez Boix}
\affiliation{
	\institution{Vrije Universiteit Brussel}
	\city{Brussel}
	\country{Belgium}
}
\email{egonzale@vub.be}

\author{Hanspeter M\"{o}ssenb\"{o}ck}
\affiliation{
	\institution{Johannes Kepler University}
	\city{Linz}
	\country{Austria}
}
\email{hanspeter.moessenboeck@jku.at}

\renewcommand{\shortauthors}{D. Aumayr, S. Marr, E. Gonzalez Boix, H. M\"{o}ssenb\"{o}ck}
\begin{abstract}

The actor model is popular for many types of server applications.
Efficient snapshotting of applications 
is crucial in the deployment of pre-initialized
applications or moving running applications 
to different machines, 
e.g for debugging purposes.
A key issue is that snapshotting blocks all other operations.
In modern latency-sensitive applications,
stopping the application to persist its state
needs to be avoided, %
because users may not tolerate the increased request latency.

In order to minimize the impact of snapshotting
on request latency,
our approach persists the application's state asynchronously
by capturing partial heaps,
completing snapshots step by step.
Additionally, our solution 
is transparent
and supports arbitrary object graphs.
We prototyped our snapshotting approach on top of the Truffle/Graal platform
and evaluated it with the Savina benchmarks
and the \AcmeAir microservice application.
When performing a snapshot every thousand \AcmeAir requests,
the number of slow requests (\AcmeLatencySHundredShare of all requests) with latency above 100ms
increases by\AcmeLatencyIncreaseHundred.
Our Savina microbenchmark results detail
how different utilization patterns impact snapshotting cost.

To the best of our knowledge, this is the first system that
enables asynchronous snapshotting of actor applications,
\ie without stop-the-world synchronization,
and thereby minimizes the impact on latency.
We thus believe it enables
new deployment and debugging options for actor systems.

\end{abstract}

\ccsdesc[500]{Software and its engineering~Software testing and debugging}
\ccsdesc[300]{Theory of computation~Concurrency}

\keywords{Actors, Snapshots, Micro services, Latency}

\maketitle

\section{Introduction}  %
Snapshotting
persists
a program's state
so that the program can be restored
and continued later.
Programming environments
such as Lisp and Smalltalk
use snapshotting
to create images of 
the system's state.
These images allow developers
to deploy a pre-configured system
into production
or continue development 
at a previous state.
Snapshots can also facilitate
time-traveling
and record\,\&\,replay debugging\citep{barr:2016:time, Barr:2014:TAT},
by restoring a program execution
to an earlier
point in time
to investigate
events that lead
to the occurrence
of a bug.
Finally, snapshots enable 
quick crash recovery
and moving a program's execution
to a different machine.

This paper focuses on snapshotting support for actor-based applications.
Popular implementations
of the actor model
such as \citeurl{Akka,}{Akka Website}{}{}{https://akka.io/}
Pony\citep{Clebsch:2015:DCS}, Erlang\citep{Erlang}, 
Elixir\citep{Thomas:2014:ELIXIR}, and
Orleans\citep{Bykov:2011:OCC}
are used to build complex responsive applications.
We present a novel technique
to snapshot
such responsive actor-based applications
avoiding stop-the-world pauses.%

Creating a snapshot
requires the program state
to be persisted.
In Lisp, Smalltalk, and other systems\citep{Barr:2014:TAT}, this is done
with heap dumps
integrated with the garbage collector (GC).
This, however,
requires virtual machine (VM) support
and usually \emph{stops the world}
to create a consistent snapshot,
making the program unresponsive.
This is problematic for applications that aim to respond consistently with low latency.

Snapshotting is also common
in high-performance computing\citep{Chandy:1985:DSD:214451.214456,Elnozahy:2002:CheckpointSurvey,BUNTINAS200873,erb2017consistent,LOSADA:2019:Relaxed} to address distributed failures.
These approaches provide inspiration
but in this work 
we do not address such failures
and focus on non-distributed actor applications, 
simplifying the problem significantly.

In this paper, we present
an efficient approach
for transparent asynchronous
snapshots of
non-distributed
actor-based systems.
It is implemented on top of an 
unmodified Java VM
and does not rely on
GC integration. %
By snapshotting the state
of each actor
individually,
we avoid
stop-the-world
synchronization that blocks
the entire application.
We applied our approach
to communicating event loop (CEL)
actors\citep{Miller:2005:CSP} in \SOMns.
\SOMns is an implementation
of Newspeak\citep{Bracha:2010:NS}
built on
the Truffle framework
and the Graal
just-in-time compiler\citep{Wurthinger:2017:PPE}.
We evaluated
the performance
of our approach
with the \SOMns implementation.
On the Savina benchmark suite\citep{Imam:2014:SAB},
we measure
the run-time overhead and memory impact
of frequent
snapshot creation.
On the modern
web-application \AcmeAir\citep{Ueda:2016:AcmeAir},
we measure the effect
of snapshot creation
on request latency,
to ensure that
user experience
remains acceptable,
\ie additional latency is below 500ms\citep{arapakis2014impact}.

The main contribution of this paper
is a novel snapshotting approach
for non-distributed actor programs
that minimizes latency
by avoiding stop-the-world synchronization
and persisting program state asynchronously.
Furthermore, our approach does not require
changes to the VM nor GC integration. %
Our evaluation shows that for the Acme Air experiment
snapshotting increases the number of slow requests 
with latency over 100ms
by\AcmeLatencyIncreaseHundred
while the maximum latency is unchanged.

\section{Background and Requirements for Asynchronous Actor Snapshots} %
\label{sec:background-motivation}

This section provides
the background
on actor concurrency
to show the challenges
of designing
an asynchronous snapshot
mechanism for
actor systems.
We also briefly discuss \SOMns, 
a programming language
with communicating event loop actors,
for which we implemented 
our approach.

\subsection{Communicating Event Loops (CELs)} %
\label{sec:CEL}

Originally, the actor model
was proposed by \citet{ActorsFormalism}.
Since then
it has been
used as an inspiration
for many
derived variants\citep{DeKoster:2016:YAT}.
In this work,
we focus
on programs written in the communicating event loop (CEL)
variant,
which was first
described for the programming language E\citep{Miller:2005:CSP}.
This variant has all the characteristics
typically associated with actor models:
isolation of state and message passing.
Additionally, it provides
non-blocking promises
as a high-level abstraction
for returning results
of asynchronous computations.
CELs were also adopted by languages
such as AmbientTalk\citep{VanCutsem:2012:AMA}
and Newspeak\citep{Bracha:2010:NS},
and correspond
to the event loops
in widely used systems
such as JavaScript
and Node.js.

\begin{figure}
	\centering
	\includegraphics{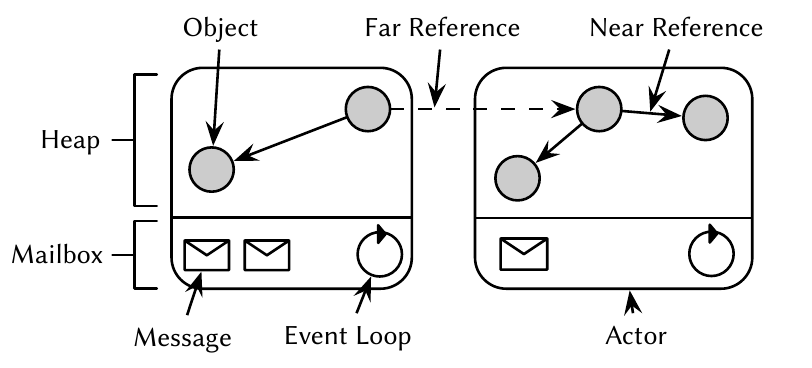}
	\caption{Main components of communicating event loops.}
	\label{fig:CEL}
\end{figure}

The structure
and main components
of CELs are shown in \cref{fig:CEL}.
Each actor contains
a set of objects
that are isolated
from those of other actors,
a mailbox for receiving messages,
and an event loop.
The event loop
of an actor
perpetually takes messages
from the mailbox
in the order of their arrival
and processes them
one-by-one.
Each message specifies
a receiver, \ie an object of the actor which understands the message,
and executes
the corresponding method
defined in the object.
The processing of a message
is an atomic operation
with regard to other messages
on the same actor
and defines a so-called \emph{turn}.

To guarantee state isolation,
each actor can only
access its own objects
directly.
Objects in other actors
are accessed via \emph{far references},
which restrict interactions to asynchronous messages.
Objects given as parameters in messages are by default passed by far reference
to ensure isolation of state.

Asynchronous message sends
immediately return
\sloppy{
a \emph{promise}}
(also known as \emph{future}).
The promise
is a placeholder
for the result
of the message
send.
The receiver of the message
promises to provide
the result
at a later time.
When the result becomes available
the promise is \emph{resolved}.
Promises are objects themselves
and consequently can receive messages.
A promise handles messages
differently from other objects:
instead of executing a method,
received messages are accumulated
in the promise while it is not resolved.
When a promise with stored messages
is resolved, all the messages
are forwarded
to the final value
of the promise.
It is also possible
to resolve a promise
with another promise 
(\ie \emph{promise chaining}).
In this case,
the resolved
promise is registered
as a dependent
in the other promise.
The promise
is fully resolved
when the other promise
is resolved,
and only then
forwards stored messages.
Similar to passing objects by far reference when they cross actor boundaries,
promise chaining
is used
when promises
are passed
to other actors.
For the receiver
of the message,
a new promise is created
that is resolved
with the original.

\begin{figure}
		\centering
		\includegraphics{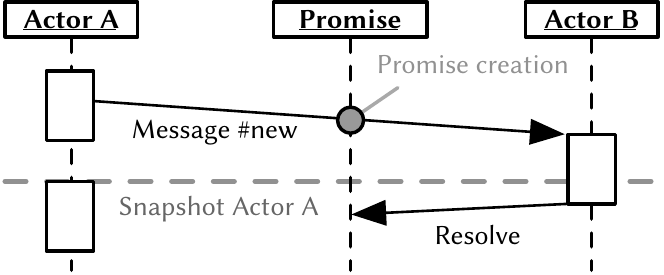}
		\caption{Race condition between serializing \code{Promise} as part of \code{Actor A}, and its resolution by \code{Actor B}.
			In an optimized implementation,
			the promise value may be lost.}
		\label{fig:promiserace}
\end{figure}

\paragraph{\SOMns}

We implement our snapshotting approach for actor programs in \SOMns, 
an implementation
of Newspeak\citep{Bracha:2010:NS}
built on
the Truffle framework
and the Graal
just-in-time compiler\citep{Wurthinger:2017:PPE}.
\SOMns implements
a CEL actor model (cf. \cref{sec:CEL})
and is designed
for shared-memory
multicore systems.
Shared memory
is used for
optimizations
where applicable.
For instance,
far references
are placeholder objects
that contain direct pointers
to objects in other actors.
\SOMns does not expose
or modify the underlying GC and VM platform.  
As such, it can be used
on top of a stock JVM
with the JVM compiler interface (JVMCI).

\subsection{Asynchronous Snapshotting for Actor Systems} %
\label{sec:problemstatement}

As described in the introduction,
our goal is to create
transparent asynchronous snapshots
for actor programs
with isolated state.
In the following, we 
describe the issues 
that have to be considered
in the design of a snapshotting solution
for CELs:
(1) serialization of far references, (2) lost promise resolution,
and (3) serialization of messages. %
While the serialization of far references and lost promise resolution
are specific to the CEL model,
the serialization of messages is required for
any actor model variant.

\paragraph{Issue 1: Serialization of Far References}
\label{issue1}
As mentioned before,
objects are passed by far reference
to maintain state isolation, 
making all
intra-actor
communication asynchronous.
However, serializing objects by
naively traversing
the object graph
of an actor
may reach
state that belongs
to another actor via far references.
Since the other actor may concurrently change its state,
snapshotting needs to account for it.
Specifically, there are two issues to tackle:
(1) Data races may occur
when the owner
of an object reached via far reference
is concurrently processing
a message and changes
its state,
when a snapshot is recorded.
This can lead to
inconsistent data
in the snapshot.
(2) Would the traversal
not stop
in the referenced actors,
but continue with
the far references
present in those as well,
the entire
program state
may be potentially serialized
at once. 
This would increase latency,
and thus, defeat the purpose
of asynchronous snapshots.
\paragraph{Issue 2: Lost Promise Resolutions}
Recall from \cref{sec:CEL}
that promises
are used
to asynchronously
return results
of message sends.
Different implementation strategies
can be used
to resolve promises.
The classic approach is to resolve them with an explicit asynchronous message
that is handled as any other message, avoiding race conditions.
Alternatively, to minimize the number
of messages sent,
promises can also be
resolved directly using locking at the implementation level,
\eg in \SOMns.
This conceptually violates
the state isolation between actors,
as one actor
can change
the state of
a promise
owned by another actor.
But, in practice, this
data race cannot be observed by a program,
because the state (and value)
of a promise
can only be accessed
asynchronously
by sending a message.
However, a snapshotting approach that is applicable
to optimized actor systems, such as \SOMns,
needs to take special care with promise resolutions as 
this data race could lead to
inconsistent or incomplete snapshots.

\Cref{fig:promiserace} shows a scenario
where the resolution of a promise is lost,
because the snapshot
takes place after
the actor's state was serialized. %
This particular scheduling
serializes an unresolved \code{Promise} object
which is owned by \code{Actor\,A}.
The promise is then resolved with an object belonging to \code{Actor\,B},
but the snapshots do not reflect that promise resolution, because there are no promise resolution messages that could be captured.
Hence,
we refer to such
a 
scenario
as a \emph{lost promise resolution}.
Its effect is that
after snapshot restoration,
the promise
is unresolved,
and since the resolving message
was not part of the snapshot,
it remains unresolved indefinitely.
Any messages
sent to the \code{Promise} object
after the snapshot is restored
would accumulate and never be delivered.

\paragraph{Issue 3: Serialization of Messages}
To successfully restore
a snapshot,
it has to contain
the messages
that were about to
be executed by an actor,
\ie, the mailbox contents.
In the context
of asynchronous 
non-blocking
snapshots
this is a challenge
as actors finish processing
their current messages
before snapshotting.
For some time
both snapshotted and un-snapshotted
actors may coexist.
Snapshotted actors
may receive
additional messages
from un-snapshotted actors,
and vice-versa.
A snapshotting approach
therefore
has to be able
to recognize messages
that were sent
from un-snapshotted
actors to snapshotted
actors and add
them to the snapshot.

\section{Snapshotting Actor Systems} %

We now present our snapshotting approach, which is designed for
actor systems
with isolated state and 
atomic message execution. %
The main idea of our approach is to perform 
snapshotting asynchronously to avoid stop-the-world
pauses and minimize application latency.
Furthermore, the snapshotting mechanism is designed
to be transparent to programs,
\ie, can be used without %
annotating code to specify
state that needs to be captured %
and to support arbitrary object graphs.
Since the actor model
ensures that
objects owned
by an actor
remain unchanged
when the actor is
not processing
a message,
our approach captures
the state
of actors
before they start
processing messages
after a snapshot
was triggered.
In addition to the actor's state,
we persist
all unprocessed messages 
that originate from
before the snapshot.
When restoring a snapshot,
the recorded messages are used to
restore an actor's mailbox
in the correct order.
In this section we detail how to capture snapshots and \cref{sec:restoringSnapshots} provides details on restoring snapshots.

\subsection{Capturing Snapshots}
\label{sec:capturingSnapshots}

To minimize latency when snapshotting an application,
we designed an asynchronous snapshotting mechanism
that is based on capturing partial state for each actor (cf. \cref{sec:partialHeaps}).
Capturing partial state is possible due to
the isolation of state and atomic processing of messages in actor programs.
To ensure completeness of the snapshots, we capture the following components of an actor's state:
(1) state directly owned
by an actor, (2) messages and near-referenced objects reachable
from them (cf. \cref{sec:partialHeaps}), 
and (3) objects that are far-referenced 
from other actors (cf. \cref{sec:deferredSerialization}).

Snapshotting
can be triggered asynchronously
either explicitly
by user code
or automatically.
Automatic snapshots can be created for example
at regular
time intervals
or when used 
alongside record \& replay (cf. \cref{sec:replayIntegration})
after the trace
size reaches a
defined limit. 
For simplicity our approach creates only one snapshot at a time, \ie snapshot creation has to be completed before another snapshot can be initiated.

Before an actor starts processing the first message after a snapshot was triggered,
we persist
its directly
owned state.
As stated in
\emph{issue 3} (cf. \cref{sec:problemstatement}),
we also need to 
identify messages
that were sent before
the snapshot request
and to serialize them
whenever they are
about to be processed
by an actor.
We solve this issue
by dividing the
program execution
into phases.
Each time
a snapshot is triggered,
a new phase begins
and a global phase counter
is incremented.
When a message is sent,
we attach the phase number 
of the sending actor to it.
Hence,
finding out if a message
needs to be captured
is a simple
numeric comparison.
All messages
that are about
to be processed
and whose send phase
does not match
the global phase
have to be serialized.
Note that actors only change
phases at the beginning of a turn, i.e., 
when they start
processing a message.
Because
the processing of messages
is atomic,
actors that are busy executing a message
cannot immediately
observe a change of 
the global phase counter
and remain in the phase
in which they started
processing their
current message
until it is completed.
This means that
such actors may send messages
with phase numbers
smaller than
the current global phase.

\begin{figure}
	\centering
	\includegraphics{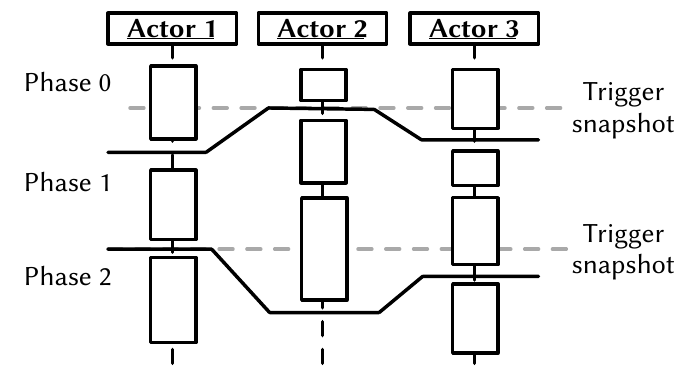}
	\caption{
		Asynchronous snapshots divide a program execution into different phases.
    Since actors only switch phases when they are not executing a message,
    phases can overlap.%
	}
	\label{fig:snapphases}
\end{figure}

\Cref{fig:snapphases} illustrates a program execution with snapshots.
It shows
that different actors
may be in different phases,
\emph{Actor 1} is the last actor
to finish processing
of its current message
and to advance to \emph{Phase 1}.
Any messages sent
by \emph{Actor 1}
after the snapshot
but before finishing
the message
are part of \emph{Phase 0}
and have to be captured
by the receiving actor.
A snapshot is said to be complete
when there are no more
actors in the previous phase,
\ie, no more messages have to be captured.
As long as actors
are in the previous phase,
additional messages may need
to be captured by actors
that already advanced.
We assume that actors
finish processing messages
and do not remain in one phase indefinitely.
Otherwise, only state which is far referenced
by other actors will be part
of the snapshot.

As messages are assigned phase numbers
when they are originally sent, 
we have to update the phase numbers
of messages when they are 
forwarded to the result of a promise resolution.
Hence, these messages will not be captured
for the current snapshot,
but are reproduced in the restored execution
when the promises containing them are resolved.

\subsection{System Architecture}%
\label{sec:architecture}

\begin{figure}
	\centering
	\includegraphics{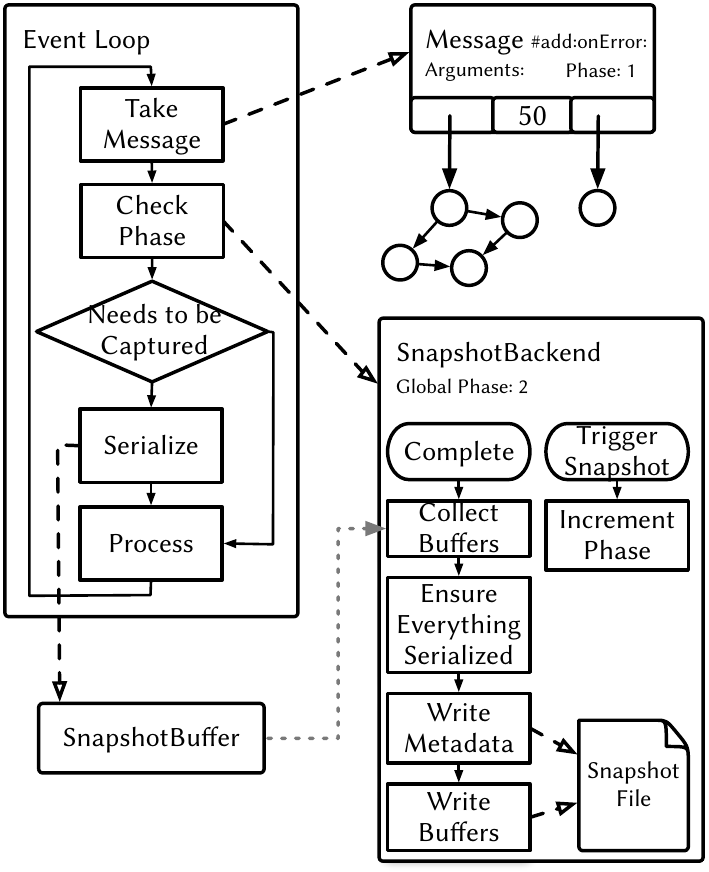}
	\caption{
		Architecture of our 
		solution showing
		how the components connect
		to produce snapshots.
    }
	\label{fig:architecture}
\end{figure}

We now describe 
the architecture of our
snapshotting approach which 
is shown in \cref{fig:architecture}.
When an event loop is being snapshotted, 
it will still take a message out of its mailbox but then
the event loop needs
to perform additional steps
before processing a message.
In the first step,
the event loop checks the phase
of the message
against the global phase
in the snapshot backend
and determines
if the message
needs to be serialized.
If that is the case,
the message is then serialized into a 
snapshot buffer.
Snapshot buffers
store the data as byte arrays in memory.

As an optimization,
in \SOMns the actors
are scheduled
flexibly
on a thread pool.
This can be used
to reduce the number
of buffers needed.
Instead of assigning
a snapshot buffer
to each actor,
buffers can be assigned 
to the processing threads.
Each buffer is
used by only one actor
at a time,
but may contain data
of different actors.
As actors can be
executed on different
processing threads,
their snapshot data
can be spread
across multiple buffers.

The snapshot backend
is 
responsible
for triggering snapshots,
maintaining metadata,
collecting the snapshot buffers
from the threads,
and writing the snapshot files.
Snapshots are written to disk by a separate
thread once they are complete.
We detect if snapshots are complete
by queuing a special task
in the thread pool used for actors.
After the task is executed,
there are no more actors
with messages that need to be captured
in the thread pool's queue.
Currently executing actors
may still contain messages
from the previous phase.
Hence, we have to wait for
each thread to finish processing it's current task,
\ie actor, before the snapshot can be persisted to disk.
At this point, there may still be some left-over deferred serializations (c.f. \cref{sec:deferredSerialization})
that need to be handled before the snapshot can be persisted.
We have bookkeeping in place that tells us which actors
have unfinished deferred serializations,
and proceed by scheduling all those actors for execution
so that they may serialize missing objects.
When all of them are done, we check if this caused some new deferred serializations,
and if yes repeat the scheduling and checking.
Only after there are no more messages to capture 
and all deferred serializations have been handled, 
we can persist the snapshot to disk.

If the root cause of an error
is before the latest snapshot,
earlier snapshots have to be used.
Otherwise, only the effects
can be reproduced and observed.

Compared to a thread-based concurrency model,
we rely on CEL actor turns to terminate.
This means, we assume that turns do not contain infinite loops.
Infinite loops are generally considered a bug in CEL actor programs,
since they impact latency and performance more generally.  
In a thread-based system,
such an infinite loop
could continuously add elements
to a data structure,
which would make it difficult to
persist the data structure.

\subsection{Capturing Partial Heaps} %
\label{sec:partialHeaps}

\begin{figure}
	\centering
	\begin{subfigure}{.27\linewidth}
		\centering
		\includegraphics{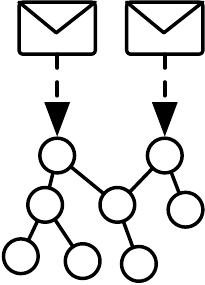}
		\subcaption{
    		The state of \emph{Actor 3} at the snapshot.
    	}
		\label{fig:snapstate}
	\end{subfigure}
	\hfill
	\begin{subfigure}{.67\linewidth}
		\centering
		\includegraphics{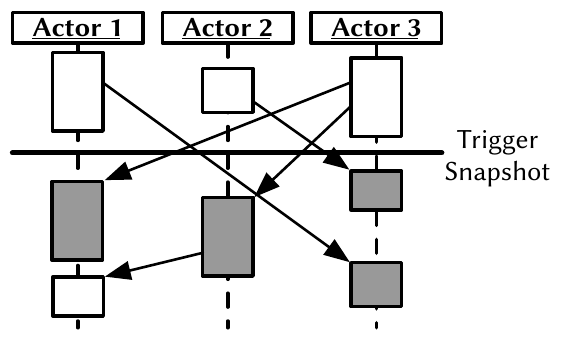}
		\subcaption{
			Execution of an actor program around the time of a snapshot.
			Messages marked grey were sent before the snapshot.
		}
		\label{fig:snapsequence}
	\end{subfigure}
	\caption{Example where partial heaps can be snapshot.}
\end{figure}

The actor state, \ie, 
the object graph
that can be reached
from different messages
in the mailbox of an actor  %
is typically partial.
For some actors, the whole object heap can be reached via a single %
message,
while for others state may consist of multiple disjoint or partially-connected object graphs.
The latter is depicted in \cref{fig:snapstate}.
To minimize the impact on message latency,
we capture only partial
state per message if possible.
The idea is to
first persist the state
a message depends on
and to process that message
afterwards.
This is done incrementally
for subsequent messages,
\ie before processing
another message, we capture
the state that was not persisted before.
Splitting the serialization %
enables actors
to make progress
and be more responsive,
as they gradually
complete the snapshot.
State that is only reachable from far references
is also captured as we explain in \cref{sec:deferredSerialization}.

In the example in \cref{fig:snapsequence},
two messages
(marked grey)
were sent
to \emph{Actor 3}
before a snapshot.
As these messages
are processed
after the snapshot,
they need to be captured.
The state of \emph{Actor 3}
that is reachable from these messages
can be seen in \cref{fig:snapstate}.
Different parts
of the actor's state
can be reached
from each of them,
but they also have
a common subgraph.
This means
that the state
of the actor
does not have
to be captured
all at once,
instead we
can capture
the state incrementally
message by message,
each message
completing the snapshot
with the reachable state
that was not captured before.
This incremental capturing
of the state
can increase
the responsiveness
of actors
after a snapshot
was triggered,
but only if there is an overlap 
in the object graph reachable from the messages, 
as in \cref{fig:snapstate}.

\subsection{Far References -- Deferring Serialization}
\label{sec:deferredSerialization}
As described
in the problem statement (\cref{issue1}),
far references
connect
the object graphs
of different actors.
To ensure correctness, \ie prevent data races,
far-referenced objects have to be serialized
by their owner.
We call this \emph{deferred serialization}.
As our solution is asynchronous,
we cannot wait for another actor
to serialize a far-referenced object
and return its location.
In addition to the far-referenced object,
the owner is handed
a buffer plus offset
where the reference information
has to be filled in
after serialization.
If the object
was already serialized,
it is not added to the
queue of deferred serializations.
Instead the reference
information is written
to the buffer immediately.
Before an actor
starts processing
a message
it handles all the
serializations
that were deferred
to it.
As actors can be idle
for longer
periods of time,
we have to consider
the possibility
that an actor
has deferred serializations
when a snapshot is written to disk.
To avoid having
missing pieces
in our snapshot
we keep track
of actors
with deferred serializations.
We force those actors to perform 
the deferred serializations
until the snapshot is complete,
\ie there are no more deferred serializations.
Hence,
even actors
that were not active
between triggering a snapshot
and writing it to disk
are captured
and can be used
after restoring the snapshot.
We achieve forced serialization
by scheduling all the actors
with deferred serializations
on the thread pool we use for execution,
even if they do not have any messages to process.
This avoids data races that
could happen if the forced serialization
was, for example, done in the snapshot writer thread.

\subsection{Promises}
\label{sec:promiseSerialization}
Promises require special
treatment when they
are serialized
as they
can contain
an arbitrary number
of messages
and references to
other promises (cf. \cref{sec:CEL}).
The referenced promises
typically belong
to a different
actor.
We therefore
defer the serialization
of those promises
to the respective owners
as explained above (cf. \cref{sec:deferredSerialization}).

Promises are local to their owner.
When sent to another actor, a new promise is created for the receiver
and chained
with the original one (cf. \cref{sec:CEL}).
Thus, all messages
that a promise receives
were sent by its owner,
\ie
there is no race
regarding the messages
contained in a promise.

Note that when serializing
a promise,
it can be
either a promise
that was already
resolved,
or an unresolved one.
The messages referenced by an unresolved promise can be serialized like any other object
since they are owned by the actor that owns the corresponding promise.
Messages related to resolved promises were already sent to another actor
and are instead 
captured by the receiver if necessary.

As discussed in \cref{sec:problemstatement},
we need to handle promise resolutions that race with the snapshotting,
which otherwise could get lost.
This could occur when a promise is serialized as unresolved,
but is then resolved by an actor in a previous phase,
which would not be reflected in the snapshot.
We detect such racy resolutions
by comparing
the resolving actor's
phase number with the global one.
If they do not match,
the resolution is captured separately
in the snapshot.

\subsection{Snapshot Format} %
An important part of serialization
is to use a format that can be efficiently recorded and read,
while preserving data in the correct order.
We now briefly detail the snapshot format used and relevant correctness concerns. 

\Cref{fig:Snapshot} shows 
a representation
of our binary
snapshot format.
It consists of two kinds of data.
First,
it has a section
of metadata
with
a registry of the snapshot messages,
offsets of the snapshot buffers
within the snapshot file,
and data to
reconstruct
promise resolutions
that are otherwise lost.
The rest of the format
consists of interconnected
heaps that represent
the serialized
object graphs.

We now detail the different components of the metadata.
To ensure correct execution when loading a snapshot,
we need to restore
the snapshotted messages
to the mailboxes
of their respective actor
in the original order.
As such, for each snapshot message,
the \code{MessageRegistry}
contains the location
where the message
object is stored
in the snapshot.
To ensure that the messages
are restored
correctly,
the registry
also contains the id
of the actor,
and an ordering number.

Promise resolutions are also
part of the metadata
and can be found as \code{Resolutions}.
This part links unresolved promises
to a result
and identifies
the actor
that performed
the resolution in
the original execution
as well as
whether it was successful or erroneous. 
\Cref{sec:promiseSerialization}
details why
we need to capture
promise resolutions
separately.

The \code{ClassEnclosures}
are used to support nested classes.
This is necessary to support the Newspeak semantics\citep{Bracha:2010:NS}
but would apply to other languages such as Java,
too.
It contains a mapping
of class ids
to the object
that enclosed
the class
in the original
execution.

The final piece
of metadata
is the \code{HeapMap};
it is a mapping of
buffer ids to offsets
in the snapshot file.
\code{HeapMap} is used to
decode our reference representations.
Each reference
consists of a 16-bit
buffer id, and a 48-bit
offset.
Referenced objects can
be found in the snapshot file
by getting the start location
of the containing buffer
and then adding the offset.

\begin{lstlisting}[caption={EBNF grammar for our binary snapshot format.},captionpos=b, label=fig:Snapshot, basicstyle=\small\ttfamily]
Snapshot = MetaData Heaps.
MetaData = MessageRegistry Resolutions 
ClassEnclosures HeapMap.
MessageRegistry = msgCnt {actorId msgNo 
msgLocation}.
ClassEnlosures = cnt {classId outerObject}.
Resolutions = cnt {resolver result actor 
state}.
HeapMap = nHeaps {heapId offset}.
Heaps = {Object}.
Object = classId ObjectData.
ObjectData = Message Promise Array ...
\end{lstlisting}

\section{Restoring Snapshots}
\label{sec:restoringSnapshots}

Restoring snapshots boils down to
deserializing
all captured objects, recreating actors,
and initializing the mailboxes
with the snapshot messages
in the right order.
The first step
in the deserialization
is to parse
the \code{MetaData}
section of the snapshot,
which contains
the storage locations
of the messages
and the start locations
of the different heaps
in the snapshot file.
Afterwards, the messages can be deserialized using the \code{MessageLocations}.
The messages
are then put
into the mailboxes
of the respective
actors
in the same order
they are in the snapshot,
which ensures they are processed
in the original execution order.
For simplicity,
actors are not allowed
to process messages
until after
all messages
are deserialized
and in a mailbox.
Otherwise
the actors would start
sending each other messages
and altering their state
while we are still restoring the program state.

Note that a snapshot
does not contain
a list of actors.
Instead, actors are created implicitly
when their ids
are encountered
during snapshot parsing.

\subsection{Deserializing Messages and Object Graphs}
We now provide further details on the deserialization.
Messages,
like any other
object in the snapshot,
are encoded
as seen in \cref{fig:Snapshot}, %
and their entry
starts with a \code{classId}.
The deserialization
of an object
starts by looking up
the class with that id.
Previously unknown
classes are loaded lazily,
\ie, the first time
they are encountered
during deserialization.
The looked-up class
allows us to deserialize
its instances transparently,
similar to serialization during snapshot-creation,
and transitively deserializes
any other referenced objects.

As different messages
may reference
the same object,
and the object graphs may have circles,
we keep track
of which objects
were already deserialized.
In the case of cyclic
object graphs,
we omit the cyclic reference
and fix it later
when both objects are
available.
Our implementation
uses a map
of an object's snapshot address
to its instance.
We can therefore
check if an object
was already deserialized
and can maintain object identity
by using the map entry
instead of deserializing it again.
For handling cycles,
we install a placeholder
in the map
before deserialization
of an object.
When during deserialization a reference is encountered
for which a placeholder exists,
instead of deserializing it
and causing an
infinite loop,
we leave the reference uninitialized
and add
some fixup information
to the placeholder.
This can be, for example,
a tuple of the object
and the field
that needs to be fixed.
Finally,
when a placeholder
is replaced
by the actual object,
all the fixups
that accumulated
are performed,
\ie, references to the object
are initialized.

\section{Evaluation} %

This section evaluates
the performance
of our snapshotting approach.
Since \SOMns is a research language,
we first compare its performance to other language implementations
to provide sufficient context.
We measure the run-time overhead and memory impact of snapshotting using
the Savina\citep{Imam:2014:SAB} benchmark suite.
We assess the impact on request latency
with the \AcmeAir microservice application,
Following \citet{arapakis2014impact},
snapshots should not increase request latency by more than 500ms,
otherwise the delay will be noticed by users.
In short, we will assess \SOMns' baseline performance,
the snapshot overhead on microbenchmarks,
and their impact on a microservice application.

\subsection{Methodology}
\SOMns relies on 
dynamic compilation 
to reach its peak performance.
Thus, to account for the VM's
warmup behavior\citep{Barrett:2017:VMW},
we executed each of
the Savina and
Are We Fast Yet\citep{Marr:2016:AWFY} benchmarks
for 1000 iterations within
a single process.
Manual inspection of the complete run-time plots
indicates that the performance
of the benchmarks stabilizes
after 100 iterations (cf. \Cref{app:savina} for plots showing these first 100 iterations).
We thus discarded the
first 100 iterations
to discount warmup and be
more representative for longer 
running applications.

For \AcmeAir, we use JMeter\citep{halili2008apache} 
to produce a predefined workload of HTTP requests.
The workload was defined by the Node.js version of \AcmeAir.
JMeter is configured to use 8 threads for making requests and
executes a mix of about 2 million randomly generated requests
based on the predefined workload pattern.
After inspecting the latency plots, we discarded the first 100,000 requests
to exclude warmup.

The Savina and Are We Fast Yet benchmarks were executed on a machine with
two quad-core Intel Xeons E5520, 2.26 GHz with 8 GB RAM,
Ubuntu Linux with kernel 4.4, Java 8.171 and Graal version 1.05.
\AcmeAir experiments were executed on a machine with a four-core Intel Core i7-4770HQ CPU,
2.2 GHz, with 16 GB RAM, a 256 GB SSD, MacOS Mojave (10.14.3), Java 8.151 and Graal version 0.41. 

\subsection{Baseline Performance of \SOMns}
\label{sec:baseline}

\begin{figure*}
	\begin{minipage}{.28\textwidth}
		\centering
		\AwfyBaseline{}
		\captionof{figure}{Boxplot comparing the performance of \SOMns other languages, showing \SOMns performs similar to Node.js.}
		\label{fig:awfy-baseline}
	\end{minipage}
	\hspace{5mm}
	\begin{minipage}{.67\textwidth}
		\centering
		\SavinaBaseline{}
		\captionof{figure}{Boxplot comparing the performance of Savina benchmarks in different actor languages for different numbers of cores. It shows that the performance of \SOMns is comparable to other actor implementations.}
		\label{fig:savina-baseline}
	\end{minipage}
\end{figure*}

To show
that the baseline performance
of \SOMns is competitive
with other language
implementations,
we evaluate the
sequential performance
of \SOMns, Node.js, and Java
with the Are We Fast Yet
benchmarks.
The results shown in \cref{fig:awfy-baseline}
indicate that \SOMns's
performance is at the level of Node.js,
a similar dynamic language,
but is not as fast as Java.

To show that the
CEL implementation
of \SOMns
has a performance
that is
similar to other
actor implementations,
we used the Savina benchmark suite.
This benchmark suite
was originally designed
for impure actor languages
with shared memory,
such as Akka, Jetlang, and Scalaz.
Hence,
some of the benchmarks
rely on shared memory,
which is not supported in \SOMns.
As a consequence,
only 18 out of 28 benchmarks
could be ported to \SOMns.
Our results
from running
this subset of
the Savina benchmark suite
are shown in \cref{fig:savina-baseline}
and suggest
that the actor model
of \SOMns reaches
performance comparable
to other JVM-based
implementations, despite its lower sequential performance.

Considering \SOMns overall performance,
we argue that it is a suitable platform for the discussed research,
and allows us to draw conclusions about performance
that are applicable to other state-of-the-art language implementations. %

\subsection{Savina: Worst Case Cost of Creating Snapshots}

To assess our snapshotting approach for a range of actor programs,
we compare
the warmed-up execution time
of benchmark iterations
with and without snapshotting
using the Savina benchmarks.

The overhead
of running a program
with snapshots enabled
depends on
the number of snapshots,
the number of snapshot messages,
and the object graph.
Production systems
run for a long time
and trigger snapshots infrequently.
As a result,
the overall overhead
of creating snapshots
is distributed
over a larger time frame
and is averaged out.
Our Savina benchmarks,
on the other hand,
have a short run time.
The overhead of
infrequent snapshots
might get lost
in the noise,
while selecting high snapshot frequencies
automatically increases
the measured overhead.
To compare benchmark iterations
it is important that
the number of snapshots
per iteration
is constant
for each benchmark.
Hence, we decided to do trigger a snapshot every second benchmark iteration,
which can be considered a worst case scenario.
In the Savina benchmarks, workload is often generated
by sending a large number of messages (up to hundred thousands)
in a loop. If a snapshot is triggered after the generation of the workload was started,
an equally large number of messages has to be captured due to their phase number.
This means that depending on the benchmark, a large
share of the overall messages has to be snapshot, increasing overhead.
This is especially noticeable for non-computational-intensive benchmarks 
with short run time, such as \emph{Counting}, \emph{ForkJoinActorCreation} and \emph{Chameneos},
which have large numbers of messages.
Hence, the Savina benchmark suite
presents more of a worst-case
scenario for our snapshotting implementation.
However, it does not represent real-world 
latency-sensitive applications for which we use the \AcmeAir application in \cref{sub:acmeair}.

\begin{figure}
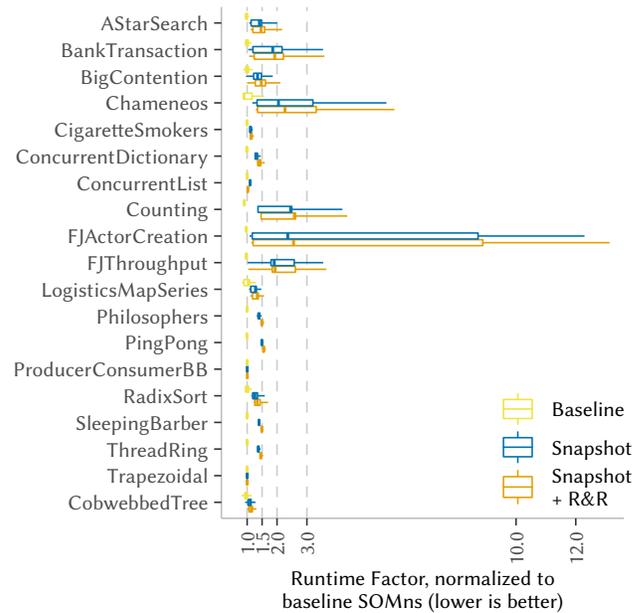

	\centering
	\SavinaSnapshots{}
	\caption{Boxplot comparing the run-time performance of Savina benchmarks when snapshotting, as well as when snapshotting is combined with record \& replay. Snapshotting performance varies depending on the benchmark, but is for most of the microbenchmarks in the 1x to 1.5x range.}
	\label{fig:savinasnapshots}
\end{figure}

Performance differences
between benchmarks
can be explained
by their different object graphs,
snapshot sizes,
and the number of messages and actors.
\Cref{fig:savinasnapshots}
shows that for snapshotting Savina benchmarks,
the run time normalized
to the mean of baseline iterations
is generally in the 1x to 1.5x area,
with a few outliers up to\SavinaMaxOverhead.
For the more computationally intensive benchmarks
of the Savina suite, such as \emph{TrapezoidalApproximation},
overhead can be as low as\SavinaMinOverheadP.

\begin{figure*}
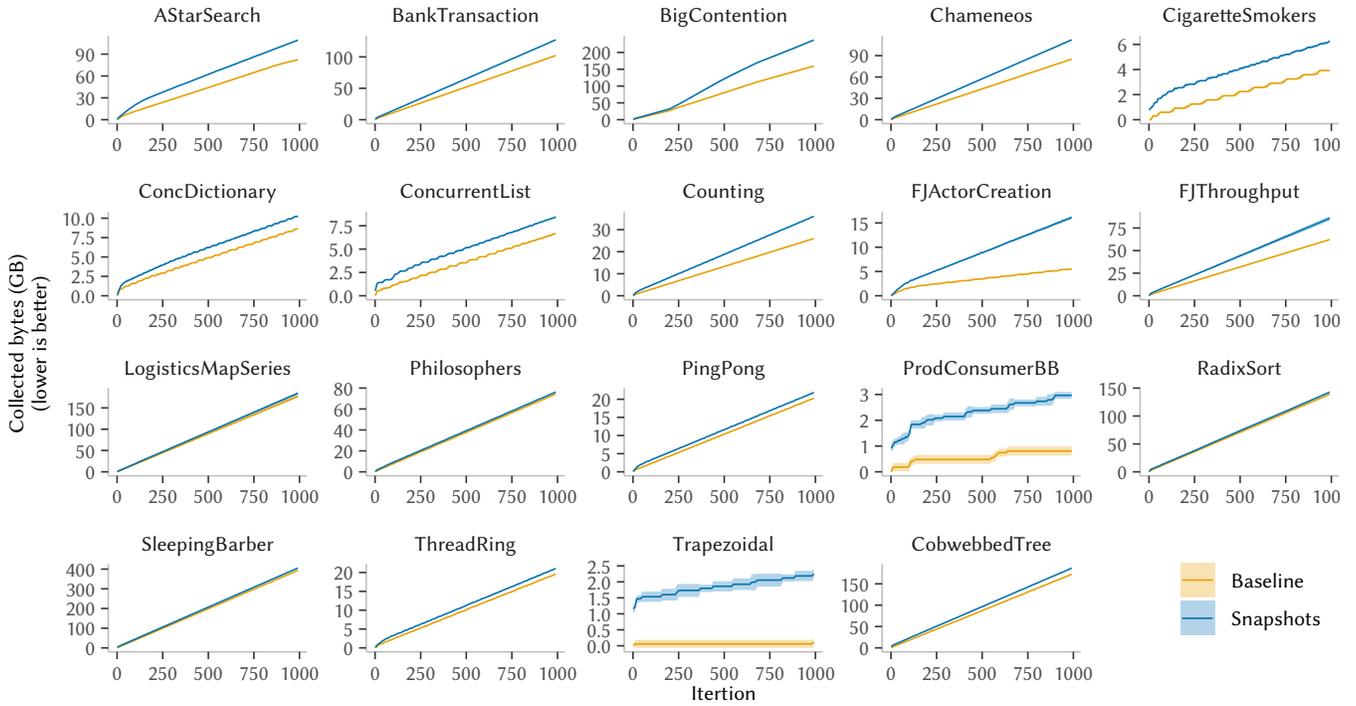

	\centering
	\SavinaCollected
	\caption{Plot of total collected bytes (GB) over 1000 iterations for the Savina benchmarks, with and without snapshotting. Most benchmarks indicate that the collected bytes increase by a small constant factor for each iteration, which is within expectations.}
	\label{fig:savinamem}
\end{figure*}

\paragraph{Memory Impact of Snapshotting Savina Benchmarks}
We also evaluated memory metrics using the Savina benchmark suite
as additional data structures and objects have an impact
on memory usage and GC behavior of an application.
We captured the metrics between benchmark iterations
to analyze their behavior over all iterations.
Due to our snapshotting, we expected to find
that the application has more and larger temporary objects, larger heap, and more time spent on GC.
\Cref{fig:savinamem} shows that the number of collected bytes is higher, and fulfills our expectations on temporary objects.
We can observe that over the course of 1000 iterations the number of 
collected bytes increases roughly linear for both configurations.
However, for snapshotting the inclination is higher, causing the number of collected bytes to diverge.
Like runtime overhead, the rate of divergence depends on the individual benchmark and ranges from close to zero (e.g. \emph{SleepingBarber}) to more than double (e.g. \emph{FJActorCreation}).
For more memory metrics we refer to appendix~\ref{app:savina}.

\paragraph{Performance of Snapshots Combined with Record \& Replay}
\label{sec:evalSavinaSnapRec}
Finally, we used the Savina benchmarks to assess how record \& replay interacts with snapshotting.
In previous work, we evaluate the performance of
record \& replay\citep{Aumayr:2018:EDR}.
Since snapshotting allows us to limit the size of recorded traces,
the two techniques should work well together.

As expected,
adding record \& replay comes at a higher
overall performance cost
than snapshotting alone.
\Cref{fig:savinasnapshots} shows that overhead increases only minimally.
The average overhead (geometric mean) increases from\SavinaAvgOverheadP with only snapshotting to\SavinaRRAvgOverheadP with additional record \& replay.

\subsection{\AcmeAir: Impact on User Experience}
\label{sub:acmeair}
User experience
is a critical factor
for the success
of responsive server applications. %
Unresponsive websites
or interfaces can discourage their use.
The latency between request and response is a key component
for the user experience.
We use \AcmeAir,
a web application
simulating the booking system
of a fictive airline,
to show the impact
of our snapshots
on the latency
of web requests.
Additionally,
we aim to show
that the maximum impact
is still acceptable.
According to \citet{arapakis2014impact},
occasional delays
of up to 500ms
are barely noticed
by users.
Thus, the impact
of our snapshots
should be
below this threshold.

For this experiment,
we decided
to trigger a snapshot
once every 1000 requests.
Every incoming request
is represented
by a message
in the actor system and
may be serialized to be part
of the snapshot.
Because \AcmeAir's
implementation uses
a cache for flights and airports
to reduce database access,
the snapshots reach
a size of up to 5MB
for the last snapshots.

\Cref{fig:acmelatency} shows the results of this experiment.
In particular, it shows the latency profile
of the different requests.
We use a logarithmic scale
for the y-axis,
so individual requests
with high latency,
for example due to GC,
are still visible.
Latency spikes,
\ie individual requests with
high latency 
did not increase dramatically
compared to baseline,
and are distributed across
the different requests.
As the different request types are not equally common,
there are some differences in the distribution of spikes.

With snapshotting enabled, the number of requests
with latency above 100ms
increases by\AcmeLatencyIncreaseHundred.
These requests make up only\AcmeLatencySHundredShare of all 20 million requests.

When looking at the slowest requests over 200ms,
we observe that the number
of such requests increases from\AcmeLatencyOriginalCntGTwohundred
to\AcmeLatencySnapCntGTwohundred requests
with snapshotting enabled.
We conclude that the latency distribution is similar to
the unmodified benchmark execution,
but there is a shift towards
higher latency,
which we attribute to the %
frequent snapshot creation.

\Cref{fig:acmebox} shows the average
latency of the
different request types.
The effect of snapshotting
is most noticeable
in \emph{QueryFlight} requests,
where the average latency
with snapshots
is up to\AcmeLatencyTracedMaxX
higher than the average latency
of the baseline.
Hence, snapshotting
has a small impact on the average
latency of requests,
while the frequenzy and latency
of the slowest requests,
which we attribute to GC,
remains similar to the original execution.
We conclude that the additional latency
for individual requests is below 500ms,
therefore our performance goal is reached
and the user experience for \AcmeAir
is not significantly impacted.

\begin{figure*}
	\begin{minipage}{.63\textwidth}
		\centering
		\AcmeTracing
		\captionof{figure}{Distribution of the total number of requests over different latencies for the \AcmeAir benchmark with snapshots every 1000 requests.
			The values with snapshotting enabled are minimally shifted.
			The number of requests over 100ms
			increases by\AcmeLatencyIncreaseHundred.
			Though, these requests make up only\AcmeLatencySHundredShare of all 20 million requests and thus seem acceptable.
		}
		\label{fig:acmelatency}
	\end{minipage}
	\hspace{5mm}
	\begin{minipage}{.33\textwidth}
		\centering
		\AcmeBox
		\captionof{figure}{Latency of \AcmeAir requests with snapshotting, normalized to baseline. Overall, latency increases by 1.66\% (geometric mean).}
		\label{fig:acmebox}
	\end{minipage}
\end{figure*}

\section{Discussion}

In our snapshotting approach, we capture a transitive closure
of all objects that are reachable from active actors and messages processed after the snapshot.
However, external resources also need to be considered for snapshots.
\SOMns supports extension modules
that can keep state, such as references to objects and actors,
which might be unreachable and therefore not part of any snapshot.
Since there is no generally correct strategy,
we leave it up to the implementers of such modules
to ensure unreachable state is persisted.
For instance, for our HTTP server as well as time-out actors,
we register their roots explicitly with the snapshotting mechanism,
so that they are recorded correctly in the snapshot.

\paragraph{Integration with Record \& Replay}
\label{sec:replayIntegration}

Our snapshotting approach
is integrated
with the \SOMns record \& replay
implementation\citep{Aumayr:2018:EDR}.
Both components can be
used individually
or together.
When combined,
the snapshot backend
notifies
the record \& replay
that a new snapshot
is created.
Any recorded trace data
afterwards
is part
of a new trace file
relative to the
new snapshot.
Old trace files
can be deleted
to free-up
disk space
if needed.
This allows 
record \& replay
to support long-running
applications by forgoing
the ability to reproduce
the execution before a snapshot.
While our snapshotting
approach captures
state, messages, and their order,
it is not a record \& replay
solution on its own.
It ensures that the captured messages are restored in order,
but does not enforce any specific interleaving with un-captured messages.
As all actors remain suspended until
the entire state is restored from a snapshot,
captured messages are not interleaved with un-captured messages,
which limits the number of observable behaviors.
In combination with record \& replay, 
one can however ensure that the order of all messages
is identical to the previously recorded one.
\paragraph{Applicability to other systems}
Our approach for asynchronous snapshotting
of actor programs should be generally applicable to
actor systems with isolated state and atomic message processing in FIFO order.
We integrated our prototype directly 
with the language implementation on top of Truffle/Graal.
However, this is not a requirement,
and instrumentation techniques, 
such as Java bytecode transformation, 
can be used as an alternative to achieve similar results.

\section{Related Work} %

In this section
we discuss related work
to provide context
for our contributions.
We compare our work
to two types
of approaches
that aim to solve
similar issues:
snapshotting solutions for distributed systems,
and back in time debugging.

\subsection{Checkpointing in Distributed Systems}
Checkpointing in distributed systems has been explored extensively.
According to \citet{Elnozahy:2002:CheckpointSurvey},
the two main approaches are coordinated and uncoordinated checkpoints.

In uncoordinated checkpointing, the distributed entities
perform checkpoints independent from each other.
When the state of one of those entities
is restored to a checkpoint, \eg due to a failure,
other entities may be forced to rollback to one of their snapshots
to keep the systems state consistent.
As the snapshots are uncoordinated, rollback
may be propagated through the system.
In the worst case, rollback is performed
until the initial state is reached.
To avoid this effect in distributed systems,
coordinated checkpointing was developed 
by \citet{Chandy:1985:DSD:214451.214456}.
The original coordinated approach is based on persisting a processes state
and then propagating marker messages when a checkpoint is created.
A process then records incoming messages for each channel, until it receives
a marker back, and adds them to the state to counteract inconsistencies.
In the following, we further discuss variants
of coordinated checkpointing.

\paragraph{Blocking vs. NonBlocking}
\citet{BUNTINAS200873} implemented and compared
blocking and non-blocking variants of coordinated checkpointing 
based on the algorithm by \citet{Chandy:1985:DSD:214451.214456}.
In the blocking variant, after checkpointing and propagating markers to all neighbors,
a process waits for all neighbors to send back a marker before execution continues.
Hence, blocking coordinated checkpoints prevent the system
from making progress until all processes have performed a checkpoint,
and consequently has a high latency.
The non-blocking variant, on the other hand, assumes that the communication channels
are guaranteed to be FIFO, and allows execution to continue
after propagating markers, while recording messages 
on incoming channels until markers are received back.
Their implementation optimized capturing
a process' state by forking, and having
the clone persist its state. This benefits
non-blocking checkpoints the most, as execution can continue
almost instantaneous.

Our snapshotting solution is non-blocking as we allow for messages
to be processed before the entire program state
is persisted. Furthermore, we do not require synchronization
between actors before they capture their state.
As we target non-distributed applications, coordination is not an issue.
We instead use a global phase number (cf. \cref{sec:capturingSnapshots}), that actors check before
they process a message. Hence, capturing state
is done lazily when an actor processes a message. %

\paragraph{Checkpoint Optimizations}
Disk-less checkpointing\citep{Plank:1998:Diskless} avoids costly disk access by 
keeping checkpoints in memory. %
or sending them over the network.
The checkpoint can be encoded and split into small chunks,
that can then be sent to a checkpoint processor, which
is responsible for recovering the state of a failed process.
Only the state of the failed process needs to be decoded,
other processes go back to the checkpoint they already have in their memory.
In contrast, our solution
keeps snapshots in memory until they are complete, 
and then writes them to disk.

Another optimization for  
distributed systems is incremental checkpointing
\citep{Naksnehaboon:2008:IncrementalCheck},
where full checkpoints are created infrequently,
with incremental checkpoints of changed memory pages captured in-between.
The incremental checkpoints are smaller and can be created faster
than full snapshots, reducing network usage when transmitting checkpoints.
Restoring incremental checkpoints
requires the last full checkpoint 
and all incremental checkpoints up to the
selected point to be processed.
For our approach to perform incremental checkpoints
we would need write barriers that update
an object's entry in the snapshot.
However, because our solution does partial heap snapshots,
we still keep latency similar to incremental approaches.

\paragraph{Coordinated Checkpointing with Relaxed Synchronization}

\citet{LOSADA:2019:Relaxed} explored coordinated checkpoint/re\-start
that enables rollback of individual processes without having to rollback others.
To achieve such a rollback, they replay messages
between restarted and non-restarted processes.
To replay messages, a message logging protocol is used to capture
non-deterministic events and messages after a process checkpoints. 
When a process is rolled back, the events in the log are replayed to bring
it to a state consistent with the other processes.
The message logging introduces additional overhead
but avoids rollback of
all processes to the last consistent state,
which compensates the overhead with faster recovery.
Our approach only captures messages
that are part of the programs state, but this can be augmented
with lightweight record \& replay to deterministically replay events
after the snapshot. However, as our solution is non-distributed,
record \& replay does not need to record message contents
as reproducing message order is sufficient.

\paragraph{Asynchronous Local Checkpointing for Actors}

To the best of our knowledge,
the only checkpointing approach for actors is for SALSA\citep{SalsaTransactors}.
Compared to our approach,
it is not transparent and it is integrated with a programming model called \emph{transactor}.
The programming model allows for some transactional behavior including rolling back actors,
which is implemented using local checkpoints.
Since we focus on snapshotting, we do not offer this kind of functionality.

\paragraph{Asynchronous Barrier Snapshotting}
Asynchronous Barrier Snapshotting (ABS) \citep{carbone2015lightweight}
is an approach for Apache Flink, 
that propagates snapshot barriers through a program.
After a process' input receives a barrier, the input is blocked until a barrier was received on all inputs,
which also causes the state of a process to be captured.
To be able to handle communication cycles, ABS relies
on static analysis to identify back-edges.
Similar to our approach, after a process receives
a barrier it does not process inputs
until its state is captured.
A big difference to our solution is in ABS's blocking behavior.
Our solution does not require actors to wait for others before
they can snapshot and continue processing messages.

\subsection{Back-in-Time Debugging}

Back-in-time debugging relies on snapshots of the program state at certain intervals, and offers time travel by replaying execution from the checkpoint before the target time.
We now compare our approach to back-in-time debugging solutions for actor-based systems.

\paragraph{Jardis} %
Jardis\citep{barr:2016:time} provides both time-travel debugging
and replay functionality for JavaScripts event loop concurrency.
It combines tracing of I/O and system calls with regular heap snapshots of the event loop.
It keeps snapshots of the last few seconds, allowing Jardis to go back as far as the oldest snapshot,
and discard trace data from before that point.
While this keeps the size of traces and snapshots small,
it limits debugging to the last seconds before a bug occurs.
When our snapshotting is combined with record \& replay, 
it provides a similar functionality
for a CEL system without relying on GC piggybacking or a modified VM.
In contrast to Jardis, our system is designed for multiple event loops
and needs to ensure correct snapshots for each event loop.

\paragraph{Event Sourcing}
Event sourcing is
a technique for actors,
where all state changes
are logged incrementally.
It
has been used
to create snapshots
of actor programs\citep{erb2017consistent}.
Their retrospective snapshots
are based on
processing an event sourcing log
and aggregating
the individual state changes
up to a certain point in time
into one independent snapshot.
This gives them the ability to
extract arbitrary snapshots
by doing post-processing.
While our solution
is designed 
to take snapshots infrequently
during program execution,
it can be combined
with \SOMns ordering-based replay\citep{Aumayr:2018:EDR}
to achieve similar results.
By restoring a snapshot
and replaying
the program execution
based on the compact trace,
we can reproduce the state
at any point after the snapshot
without having to log
all state changes
in the original execution.

\section{Conclusion and Future Work} %
The actor model has become popular
for implementation of responsive server applications.
Unfortunately, many snapshotting approaches
are blocking, and cause applications to
become unresponsive for the duration 
of snapshot creation. 

In this paper, we presented a novel
approach for creating asynchronous snapshots
of actor programs
transparently and without VM modifications or GC integration.
Our approach uses the isolation of state
of the actor model to reduce snapshot latency
by capturing partial heaps and allowing actors 
to make progress before all their state is persisted.

We evaluated the impact of our snapshotting
approach on application latency with the \AcmeAir benchmark.
Our results show that with frequent snapshotting enabled,
the latency of most requests remains below 100ms.
Only\AcmeLatencySHundredShare of 20 million requests had a latency over 100ms. 
While the number of such requests did increase by\AcmeLatencyIncreaseHundred, %
their total is still small.
We conclude that our approach
does not negatively impact the user experience
of such a web service.

\paragraph{Future Work}

As future work, we plan to apply our approach to JavaScript
with a wider range of
benchmarks.
This would also enable a direct comparison with Jardis.

\emph{Time Travel Debugging.}
While our snapshotting can already be
combined with record \& replay,
further research is needed to enable time travel debugging.
For example, restoring a snapshot currently requires
a fresh start of the program.
This is a problem as
time travel debugging requires
frequent snapshot restoration,
which means, we need to be able
to replace the whole application state within a VM.

\emph{Application Redeployment.}
To be practical for moving applications between machines,
our 
approach
has to be enhanced
so
that required resources are also moved.
This could be achieved
by
bundling the source code of all
loaded classes with the snapshot
and tracking all 
used
external resources 
so they can be made available
in the new environment.

\begin{acks}
This research is funded in part by a collaboration grant of the Austrian Science Fund (FWF) and the Research Foundation Flanders (FWO Belgium) as project I2491-N31 and G004816N.
\end{acks}

\bibliographystyle{ACM-Reference-Format}
\bibliography{paper}


\begin{thebibliography}{28}


\ifx \showCODEN    \undefined \def \showCODEN     #1{\unskip}     \fi
\ifx \showDOI      \undefined \def \showDOI       #1{#1}\fi
\ifx \showISBNx    \undefined \def \showISBNx     #1{\unskip}     \fi
\ifx \showISBNxiii \undefined \def \showISBNxiii  #1{\unskip}     \fi
\ifx \showISSN     \undefined \def \showISSN      #1{\unskip}     \fi
\ifx \showLCCN     \undefined \def \showLCCN      #1{\unskip}     \fi
\ifx \shownote     \undefined \def \shownote      #1{#1}          \fi
\ifx \showarticletitle \undefined \def \showarticletitle #1{#1}   \fi
\ifx \showURL      \undefined \def \showURL       {\relax}        \fi
\providecommand\bibfield[2]{#2}
\providecommand\bibinfo[2]{#2}
\providecommand\natexlab[1]{#1}
\providecommand\showeprint[2][]{arXiv:#2}

\bibitem[\protect\citeauthoryear{Arapakis, Bai, and Cambazoglu}{Arapakis
  et~al\mbox{.}}{2014}]%
        {arapakis2014impact}
\bibfield{author}{\bibinfo{person}{Ioannis Arapakis}, \bibinfo{person}{Xiao
  Bai}, {and} \bibinfo{person}{B~Barla Cambazoglu}.}
  \bibinfo{year}{2014}\natexlab{}.
\newblock \showarticletitle{Impact of response latency on user behavior in web
  search}. In \bibinfo{booktitle}{\emph{Proceedings of the 37th international
  ACM SIGIR conference on Research \& development in information retrieval}}.
  ACM, \bibinfo{pages}{103--112}.
\newblock


\bibitem[\protect\citeauthoryear{Armstrong, Virding, Wikstrom, and
  Williams}{Armstrong et~al\mbox{.}}{1996}]%
        {Erlang}
\bibfield{author}{\bibinfo{person}{Joe Armstrong}, \bibinfo{person}{Robert
  Virding}, \bibinfo{person}{Claes Wikstrom}, {and} \bibinfo{person}{Mike
  Williams}.} \bibinfo{year}{1996}\natexlab{}.
\newblock \bibinfo{booktitle}{\emph{Concurrent Programming in Erlang}
  (\bibinfo{edition}{2} ed.)}.
\newblock \bibinfo{publisher}{Prentice Hall PTR}.
\newblock
\showISBNx{013508301X}


\bibitem[\protect\citeauthoryear{Aumayr, Marr, B{\'e}ra, Boix, and
  M\"{o}ssenb\"{o}ck}{Aumayr et~al\mbox{.}}{2018}]%
        {Aumayr:2018:EDR}
\bibfield{author}{\bibinfo{person}{Dominik Aumayr}, \bibinfo{person}{Stefan
  Marr}, \bibinfo{person}{Cl{\'e}ment B{\'e}ra},
  \bibinfo{person}{Elisa~Gonzalez Boix}, {and} \bibinfo{person}{Hanspeter
  M\"{o}ssenb\"{o}ck}.} \bibinfo{year}{2018}\natexlab{}.
\newblock \showarticletitle{Efficient and Deterministic Record \& Replay for
  Actor Languages}. In \bibinfo{booktitle}{\emph{Proceedings of the 15th
  International Conference on Managed Languages \& Runtimes}}
  \emph{(\bibinfo{series}{ManLang '18})}. \bibinfo{publisher}{ACM}, Article
  \bibinfo{articleno}{15}, \bibinfo{numpages}{14}~pages.
\newblock
\showISBNx{978-1-4503-6424-9}
\urldef\tempurl%
\url{https://doi.org/10.1145/3237009.3237015}
\showDOI{\tempurl}


\bibitem[\protect\citeauthoryear{Barr and Marron}{Barr and Marron}{2014}]%
        {Barr:2014:TAT}
\bibfield{author}{\bibinfo{person}{Earl~T. Barr} {and} \bibinfo{person}{Mark
  Marron}.} \bibinfo{year}{2014}\natexlab{}.
\newblock \showarticletitle{Tardis: Affordable Time-travel Debugging in Managed
  Runtimes}. In \bibinfo{booktitle}{\emph{Proceedings of the 2014 ACM
  International Conference on Object Oriented Programming Systems Languages \&
  Applications}} \emph{(\bibinfo{series}{OOPSLA '14})}.
  \bibinfo{publisher}{ACM}, \bibinfo{address}{New York, NY, USA},
  \bibinfo{pages}{67--82}.
\newblock
\showISBNx{978-1-4503-2585-1}
\urldef\tempurl%
\url{https://doi.org/10.1145/2660193.2660209}
\showDOI{\tempurl}


\bibitem[\protect\citeauthoryear{Barr, Marron, Maurer, Moseley, and Seth}{Barr
  et~al\mbox{.}}{2016}]%
        {barr:2016:time}
\bibfield{author}{\bibinfo{person}{Earl~T Barr}, \bibinfo{person}{Mark Marron},
  \bibinfo{person}{Ed Maurer}, \bibinfo{person}{Dan Moseley}, {and}
  \bibinfo{person}{Gaurav Seth}.} \bibinfo{year}{2016}\natexlab{}.
\newblock \showarticletitle{{Time-travel debugging for JavaScript/Node.js}}. In
  \bibinfo{booktitle}{\emph{Proceedings of the 2016 24th ACM SIGSOFT
  International Symposium on Foundations of Software Engineering}}
  \emph{(\bibinfo{series}{FSE 2016})}. \bibinfo{publisher}{ACM},
  \bibinfo{pages}{1003--1007}.
\newblock
\urldef\tempurl%
\url{https://doi.org/10.1145/2950290.2983933}
\showDOI{\tempurl}


\bibitem[\protect\citeauthoryear{Barrett, Bolz-Tereick, Killick, Mount, and
  Tratt}{Barrett et~al\mbox{.}}{2017}]%
        {Barrett:2017:VMW}
\bibfield{author}{\bibinfo{person}{Edd Barrett},
  \bibinfo{person}{Carl~Friedrich Bolz-Tereick}, \bibinfo{person}{Rebecca
  Killick}, \bibinfo{person}{Sarah Mount}, {and} \bibinfo{person}{Laurence
  Tratt}.} \bibinfo{year}{2017}\natexlab{}.
\newblock \showarticletitle{{Virtual Machine Warmup Blows Hot and Cold}}.
\newblock \bibinfo{journal}{\emph{Proc. ACM Program. Lang.}}
  \bibinfo{volume}{1}, \bibinfo{number}{OOPSLA}, Article
  \bibinfo{articleno}{52} (\bibinfo{date}{Oct.} \bibinfo{year}{2017}),
  \bibinfo{numpages}{27}~pages.
\newblock
\showISSN{2475-1421}
\urldef\tempurl%
\url{https://doi.org/10.1145/3133876}
\showDOI{\tempurl}


\bibitem[\protect\citeauthoryear{Bracha, von~der Ahé, Bykov, Kashai, Maddox,
  and Miranda}{Bracha et~al\mbox{.}}{2010}]%
        {Bracha:2010:NS}
\bibfield{author}{\bibinfo{person}{Gilad Bracha}, \bibinfo{person}{Peter
  von~der Ahé}, \bibinfo{person}{Vassili Bykov}, \bibinfo{person}{Yaron
  Kashai}, \bibinfo{person}{William Maddox}, {and} \bibinfo{person}{Eliot
  Miranda}.} \bibinfo{year}{2010}\natexlab{}.
\newblock \showarticletitle{{Modules as Objects in Newspeak}}. In
  \bibinfo{booktitle}{\emph{ECOOP 2010 -- Object-Oriented Programming}}
  \emph{(\bibinfo{series}{Lecture Notes in Computer Science})},
  Vol.~\bibinfo{volume}{6183}. \bibinfo{publisher}{Springer},
  \bibinfo{pages}{405--428}.
\newblock
\showISBNx{978-3-642-14106-5}
\urldef\tempurl%
\url{https://doi.org/10.1007/978-3-642-14107-2_20}
\showDOI{\tempurl}


\bibitem[\protect\citeauthoryear{Buntinas, Coti, Herault, Lemarinier, Pilard,
  Rezmerita, Rodriguez, and Cappello}{Buntinas et~al\mbox{.}}{2008}]%
        {BUNTINAS200873}
\bibfield{author}{\bibinfo{person}{Darius Buntinas}, \bibinfo{person}{Camille
  Coti}, \bibinfo{person}{Thomas Herault}, \bibinfo{person}{Pierre Lemarinier},
  \bibinfo{person}{Laurence Pilard}, \bibinfo{person}{Ala Rezmerita},
  \bibinfo{person}{Eric Rodriguez}, {and} \bibinfo{person}{Franck Cappello}.}
  \bibinfo{year}{2008}\natexlab{}.
\newblock \showarticletitle{Blocking vs. non-blocking coordinated checkpointing
  for large-scale fault tolerant MPI Protocols}.
\newblock \bibinfo{journal}{\emph{Future Generation Computer Systems}}
  \bibinfo{volume}{24}, \bibinfo{number}{1} (\bibinfo{year}{2008}),
  \bibinfo{pages}{73 -- 84}.
\newblock
\showISSN{0167-739X}
\urldef\tempurl%
\url{https://doi.org/10.1016/j.future.2007.02.002}
\showDOI{\tempurl}


\bibitem[\protect\citeauthoryear{Bykov, Geller, Kliot, Larus, Pandya, and
  Thelin}{Bykov et~al\mbox{.}}{2011}]%
        {Bykov:2011:OCC}
\bibfield{author}{\bibinfo{person}{Sergey Bykov}, \bibinfo{person}{Alan
  Geller}, \bibinfo{person}{Gabriel Kliot}, \bibinfo{person}{James~R. Larus},
  \bibinfo{person}{Ravi Pandya}, {and} \bibinfo{person}{Jorgen Thelin}.}
  \bibinfo{year}{2011}\natexlab{}.
\newblock \showarticletitle{Orleans: Cloud Computing for Everyone}. In
  \bibinfo{booktitle}{\emph{Proceedings of the 2Nd ACM Symposium on Cloud
  Computing}} \emph{(\bibinfo{series}{SOCC '11})}. \bibinfo{publisher}{ACM},
  Article \bibinfo{articleno}{16}, \bibinfo{numpages}{14}~pages.
\newblock
\showISBNx{978-1-4503-0976-9}
\urldef\tempurl%
\url{https://doi.org/10.1145/2038916.2038932}
\showDOI{\tempurl}


\bibitem[\protect\citeauthoryear{Carbone, F{\'o}ra, Ewen, Haridi, and
  Tzoumas}{Carbone et~al\mbox{.}}{2015}]%
        {carbone2015lightweight}
\bibfield{author}{\bibinfo{person}{Paris Carbone}, \bibinfo{person}{Gyula
  F{\'o}ra}, \bibinfo{person}{Stephan Ewen}, \bibinfo{person}{Seif Haridi},
  {and} \bibinfo{person}{Kostas Tzoumas}.} \bibinfo{year}{2015}\natexlab{}.
\newblock \showarticletitle{Lightweight asynchronous snapshots for distributed
  dataflows}.
\newblock \bibinfo{journal}{\emph{arXiv preprint arXiv:1506.08603}}
  (\bibinfo{year}{2015}).
\newblock


\bibitem[\protect\citeauthoryear{Chandy and Lamport}{Chandy and
  Lamport}{1985}]%
        {Chandy:1985:DSD:214451.214456}
\bibfield{author}{\bibinfo{person}{K.~Mani Chandy} {and}
  \bibinfo{person}{Leslie Lamport}.} \bibinfo{year}{1985}\natexlab{}.
\newblock \showarticletitle{Distributed Snapshots: Determining Global States of
  Distributed Systems}.
\newblock \bibinfo{journal}{\emph{ACM Trans. Comput. Syst.}}
  \bibinfo{volume}{3}, \bibinfo{number}{1} (\bibinfo{date}{Feb.}
  \bibinfo{year}{1985}), \bibinfo{pages}{63--75}.
\newblock
\showISSN{0734-2071}
\urldef\tempurl%
\url{https://doi.org/10.1145/214451.214456}
\showDOI{\tempurl}


\bibitem[\protect\citeauthoryear{Clebsch, Drossopoulou, Blessing, and
  McNeil}{Clebsch et~al\mbox{.}}{2015}]%
        {Clebsch:2015:DCS}
\bibfield{author}{\bibinfo{person}{Sylvan Clebsch}, \bibinfo{person}{Sophia
  Drossopoulou}, \bibinfo{person}{Sebastian Blessing}, {and}
  \bibinfo{person}{Andy McNeil}.} \bibinfo{year}{2015}\natexlab{}.
\newblock \showarticletitle{Deny Capabilities for Safe, Fast Actors}. In
  \bibinfo{booktitle}{\emph{Proceedings of the 5th International Workshop on
  Programming Based on Actors, Agents, and Decentralized Control}}
  \emph{(\bibinfo{series}{AGERE! 2015})}. \bibinfo{publisher}{ACM},
  \bibinfo{address}{New York, NY, USA}, \bibinfo{pages}{1--12}.
\newblock
\showISBNx{978-1-4503-3901-8}
\urldef\tempurl%
\url{https://doi.org/10.1145/2824815.2824816}
\showDOI{\tempurl}


\bibitem[\protect\citeauthoryear{De~Koster, Van~Cutsem, and
  De~Meuter}{De~Koster et~al\mbox{.}}{2016}]%
        {DeKoster:2016:YAT}
\bibfield{author}{\bibinfo{person}{Joeri De~Koster}, \bibinfo{person}{Tom
  Van~Cutsem}, {and} \bibinfo{person}{Wolfgang De~Meuter}.}
  \bibinfo{year}{2016}\natexlab{}.
\newblock \showarticletitle{43 Years of Actors: A Taxonomy of Actor Models and
  Their Key Properties}. In \bibinfo{booktitle}{\emph{Proceedings of the 6th
  International Workshop on Programming Based on Actors, Agents, and
  Decentralized Control}} \emph{(\bibinfo{series}{AGERE 2016})}.
  \bibinfo{publisher}{ACM}, \bibinfo{address}{New York, NY, USA},
  \bibinfo{pages}{31--40}.
\newblock
\showISBNx{978-1-4503-4639-9}
\urldef\tempurl%
\url{https://doi.org/10.1145/3001886.3001890}
\showDOI{\tempurl}


\bibitem[\protect\citeauthoryear{Elnozahy, Alvisi, Wang, and Johnson}{Elnozahy
  et~al\mbox{.}}{2002}]%
        {Elnozahy:2002:CheckpointSurvey}
\bibfield{author}{\bibinfo{person}{E.~N.~(Mootaz) Elnozahy},
  \bibinfo{person}{Lorenzo Alvisi}, \bibinfo{person}{Yi-Min Wang}, {and}
  \bibinfo{person}{David~B. Johnson}.} \bibinfo{year}{2002}\natexlab{}.
\newblock \showarticletitle{A Survey of Rollback-recovery Protocols in
  Message-passing Systems}.
\newblock \bibinfo{journal}{\emph{ACM Comput. Surv.}} \bibinfo{volume}{34},
  \bibinfo{number}{3} (\bibinfo{date}{Sept.} \bibinfo{year}{2002}),
  \bibinfo{pages}{375--408}.
\newblock
\showISSN{0360-0300}
\urldef\tempurl%
\url{https://doi.org/10.1145/568522.568525}
\showDOI{\tempurl}


\bibitem[\protect\citeauthoryear{Erb, Mei{\ss}ner, Habiger, Pietron, and
  Kargl}{Erb et~al\mbox{.}}{2017}]%
        {erb2017consistent}
\bibfield{author}{\bibinfo{person}{Benjamin Erb}, \bibinfo{person}{Dominik
  Mei{\ss}ner}, \bibinfo{person}{Gerhard Habiger}, \bibinfo{person}{Jakob
  Pietron}, {and} \bibinfo{person}{Frank Kargl}.}
  \bibinfo{year}{2017}\natexlab{}.
\newblock \showarticletitle{Consistent retrospective snapshots in distributed
  event-sourced systems}. In \bibinfo{booktitle}{\emph{2017 International
  Conference on Networked Systems (NetSys)}}. IEEE, \bibinfo{pages}{1--8}.
\newblock


\bibitem[\protect\citeauthoryear{Halili}{Halili}{2008}]%
        {halili2008apache}
\bibfield{author}{\bibinfo{person}{Emily Halili}.}
  \bibinfo{year}{2008}\natexlab{}.
\newblock \bibinfo{booktitle}{\emph{Apache JMeter}}.
\newblock \bibinfo{publisher}{Packt Publishing}.
\newblock
\showISBNx{1847192955, 9781847192950}


\bibitem[\protect\citeauthoryear{Hewitt, Bishop, and Steiger}{Hewitt
  et~al\mbox{.}}{1973}]%
        {ActorsFormalism}
\bibfield{author}{\bibinfo{person}{Carl Hewitt}, \bibinfo{person}{Peter
  Bishop}, {and} \bibinfo{person}{Richard Steiger}.}
  \bibinfo{year}{1973}\natexlab{}.
\newblock \showarticletitle{{A Universal Modular ACTOR Formalism for Artificial
  Intelligence}}. In \bibinfo{booktitle}{\emph{IJCAI'73: Proceedings of the 3rd
  International Joint Conference on Artificial Intelligence}}.
  \bibinfo{publisher}{Morgan Kaufmann}, \bibinfo{pages}{235--245}.
\newblock


\bibitem[\protect\citeauthoryear{Imam and Sarkar}{Imam and Sarkar}{2014}]%
        {Imam:2014:SAB}
\bibfield{author}{\bibinfo{person}{Shams~M. Imam} {and} \bibinfo{person}{Vivek
  Sarkar}.} \bibinfo{year}{2014}\natexlab{}.
\newblock \showarticletitle{{Savina - An Actor Benchmark Suite: Enabling
  Empirical Evaluation of Actor Libraries}}. In
  \bibinfo{booktitle}{\emph{Proceedings of the 4th International Workshop on
  Programming Based on Actors Agents \& Decentralized Control}}
  \emph{(\bibinfo{series}{AGERE!'14})}. \bibinfo{publisher}{ACM},
  \bibinfo{pages}{67--80}.
\newblock
\showISBNx{978-1-4503-2189-1}
\urldef\tempurl%
\url{https://doi.org/10.1145/2687357.2687368}
\showDOI{\tempurl}


\bibitem[\protect\citeauthoryear{Kuang, Field, and Varela}{Kuang
  et~al\mbox{.}}{2014}]%
        {SalsaTransactors}
\bibfield{author}{\bibinfo{person}{Phillip Kuang}, \bibinfo{person}{John
  Field}, {and} \bibinfo{person}{Carlos~A. Varela}.}
  \bibinfo{year}{2014}\natexlab{}.
\newblock \showarticletitle{{Fault Tolerant Distributed Computing Using
  Asynchronous Local Checkpointing}}. In \bibinfo{booktitle}{\emph{Proceedings
  of the 4th International Workshop on Programming based on Actors Agents \&
  Decentralized Control}} \emph{(\bibinfo{series}{AGERE! '14})}.
  \bibinfo{pages}{81--93}.
\newblock
\showISBNx{978-1-4503-2189-1}
\urldef\tempurl%
\url{https://doi.org/10.1145/2687357.2687364}
\showDOI{\tempurl}


\bibitem[\protect\citeauthoryear{Losada, Bosilca, Bouteiller, González, and
  Martín}{Losada et~al\mbox{.}}{2019}]%
        {LOSADA:2019:Relaxed}
\bibfield{author}{\bibinfo{person}{Nuria Losada}, \bibinfo{person}{George
  Bosilca}, \bibinfo{person}{Aurélien Bouteiller}, \bibinfo{person}{Patricia
  González}, {and} \bibinfo{person}{María~J. Martín}.}
  \bibinfo{year}{2019}\natexlab{}.
\newblock \showarticletitle{Local rollback for resilient MPI applications with
  application-level checkpointing and message logging}.
\newblock \bibinfo{journal}{\emph{Future Generation Computer Systems}}
  \bibinfo{volume}{91} (\bibinfo{year}{2019}), \bibinfo{pages}{450 -- 464}.
\newblock
\showISSN{0167-739X}
\urldef\tempurl%
\url{https://doi.org/10.1016/j.future.2018.09.041}
\showDOI{\tempurl}


\bibitem[\protect\citeauthoryear{Marr, Daloze, and Mössenböck}{Marr
  et~al\mbox{.}}{2016}]%
        {Marr:2016:AWFY}
\bibfield{author}{\bibinfo{person}{Stefan Marr}, \bibinfo{person}{Benoit
  Daloze}, {and} \bibinfo{person}{Hanspeter Mössenböck}.}
  \bibinfo{year}{2016}\natexlab{}.
\newblock \showarticletitle{{Cross-Language Compiler Benchmarking---Are We Fast
  Yet?}}. In \bibinfo{booktitle}{\emph{Proceedings of the 12th Symposium on
  Dynamic Languages}} \emph{(\bibinfo{series}{DLS'16})}.
  \bibinfo{publisher}{ACM}, \bibinfo{pages}{120--131}.
\newblock
\showISBNx{978-1-4503-4445-6}
\urldef\tempurl%
\url{https://doi.org/10.1145/2989225.2989232}
\showDOI{\tempurl}


\bibitem[\protect\citeauthoryear{Miller, Tribble, and Shapiro}{Miller
  et~al\mbox{.}}{2005}]%
        {Miller:2005:CSP}
\bibfield{author}{\bibinfo{person}{Mark~S. Miller}, \bibinfo{person}{E.~Dean
  Tribble}, {and} \bibinfo{person}{Jonathan Shapiro}.}
  \bibinfo{year}{2005}\natexlab{}.
\newblock \showarticletitle{{Concurrency Among Strangers: Programming in E As
  Plan Coordination}}. In \bibinfo{booktitle}{\emph{Proceedings of the 1st
  International Conference on Trustworthy Global Computing}}
  \emph{(\bibinfo{series}{TGC'05})}. \bibinfo{publisher}{Springer},
  \bibinfo{pages}{195--229}.
\newblock
\showISBNx{3-540-30007-4}


\bibitem[\protect\citeauthoryear{{Naksinehaboon}, {Liu}, {Leangsuksun},
  {Nassar}, {Paun}, and {Scott}}{{Naksinehaboon} et~al\mbox{.}}{2008}]%
        {Naksnehaboon:2008:IncrementalCheck}
\bibfield{author}{\bibinfo{person}{N. {Naksinehaboon}}, \bibinfo{person}{Y.
  {Liu}}, \bibinfo{person}{C. {Leangsuksun}}, \bibinfo{person}{R. {Nassar}},
  \bibinfo{person}{M. {Paun}}, {and} \bibinfo{person}{S.~L. {Scott}}.}
  \bibinfo{year}{2008}\natexlab{}.
\newblock \showarticletitle{Reliability-Aware Approach: An Incremental
  Checkpoint/Restart Model in HPC Environments}. In
  \bibinfo{booktitle}{\emph{2008 Eighth IEEE International Symposium on Cluster
  Computing and the Grid (CCGRID)}}. \bibinfo{pages}{783--788}.
\newblock
\urldef\tempurl%
\url{https://doi.org/10.1109/CCGRID.2008.109}
\showDOI{\tempurl}


\bibitem[\protect\citeauthoryear{{Plank}, , and {Puening}}{{Plank}
  et~al\mbox{.}}{1998}]%
        {Plank:1998:Diskless}
\bibfield{author}{\bibinfo{person}{J.~S. {Plank}}, \bibinfo{person}{}, {and}
  \bibinfo{person}{M.~A. {Puening}}.} \bibinfo{year}{1998}\natexlab{}.
\newblock \showarticletitle{Diskless checkpointing}.
\newblock \bibinfo{journal}{\emph{IEEE Transactions on Parallel and Distributed
  Systems}} \bibinfo{volume}{9}, \bibinfo{number}{10} (\bibinfo{date}{Oct}
  \bibinfo{year}{1998}), \bibinfo{pages}{972--986}.
\newblock
\showISSN{1045-9219}
\urldef\tempurl%
\url{https://doi.org/10.1109/71.730527}
\showDOI{\tempurl}


\bibitem[\protect\citeauthoryear{Thomas}{Thomas}{2014}]%
        {Thomas:2014:ELIXIR}
\bibfield{author}{\bibinfo{person}{Dave Thomas}.}
  \bibinfo{year}{2014}\natexlab{}.
\newblock \bibinfo{booktitle}{\emph{Programming Elixir: Functional , Concurrent
  , Pragmatic , Fun} (\bibinfo{edition}{1st} ed.)}.
\newblock \bibinfo{publisher}{Pragmatic Bookshelf}.
\newblock
\showISBNx{1937785580, 9781937785581}


\bibitem[\protect\citeauthoryear{Ueda, Nakaike, and Ohara}{Ueda
  et~al\mbox{.}}{2016}]%
        {Ueda:2016:AcmeAir}
\bibfield{author}{\bibinfo{person}{Takanori Ueda}, \bibinfo{person}{Takuya
  Nakaike}, {and} \bibinfo{person}{Moriyoshi Ohara}.}
  \bibinfo{year}{2016}\natexlab{}.
\newblock \showarticletitle{{Workload Characterization for Microservices}}. In
  \bibinfo{booktitle}{\emph{2016 IEEE International Symposium on Workload
  Characterization}} \emph{(\bibinfo{series}{IISWC'16})}.
  \bibinfo{publisher}{IEEE}, \bibinfo{pages}{85--94}.
\newblock
\showISBNx{978-1-5090-3896-1}
\urldef\tempurl%
\url{https://doi.org/10.1109/IISWC.2016.7581269}
\showDOI{\tempurl}


\bibitem[\protect\citeauthoryear{Van~Cutsem}{Van~Cutsem}{2012}]%
        {VanCutsem:2012:AMA}
\bibfield{author}{\bibinfo{person}{Tom Van~Cutsem}.}
  \bibinfo{year}{2012}\natexlab{}.
\newblock \showarticletitle{AmbientTalk: Modern Actors for Modern Networks}. In
  \bibinfo{booktitle}{\emph{Proceedings of the 14th Workshop on Formal
  Techniques for Java-like Programs}} \emph{(\bibinfo{series}{FTfJP '12})}.
  \bibinfo{publisher}{ACM}, \bibinfo{pages}{2--2}.
\newblock
\showISBNx{978-1-4503-1272-1}
\urldef\tempurl%
\url{https://doi.org/10.1145/2318202.2318204}
\showDOI{\tempurl}


\bibitem[\protect\citeauthoryear{W\"{u}rthinger, Wimmer, Humer, W\"{o}\ss,
  Stadler, Seaton, Duboscq, Simon, and Grimmer}{W\"{u}rthinger
  et~al\mbox{.}}{2017}]%
        {Wurthinger:2017:PPE}
\bibfield{author}{\bibinfo{person}{Thomas W\"{u}rthinger},
  \bibinfo{person}{Christian Wimmer}, \bibinfo{person}{Christian Humer},
  \bibinfo{person}{Andreas W\"{o}\ss}, \bibinfo{person}{Lukas Stadler},
  \bibinfo{person}{Chris Seaton}, \bibinfo{person}{Gilles Duboscq},
  \bibinfo{person}{Doug Simon}, {and} \bibinfo{person}{Matthias Grimmer}.}
  \bibinfo{year}{2017}\natexlab{}.
\newblock \showarticletitle{{Practical Partial Evaluation for High-performance
  Dynamic Language Runtimes}}. In \bibinfo{booktitle}{\emph{Proceedings of the
  38th ACM SIGPLAN Conference on Programming Language Design and
  Implementation}} \emph{(\bibinfo{series}{PLDI'17})}.
  \bibinfo{publisher}{ACM}, \bibinfo{pages}{662--676}.
\newblock
\showISBNx{978-1-4503-4988-8}
\urldef\tempurl%
\url{https://doi.org/10.1145/3062341.3062381}
\showDOI{\tempurl}


\end{thebibliography}

\appendix

\section{Appendix}
In this section,
we present supplemental warmup and memory metrics
for the evaluation
of our snapshotting approach in the
Savina benchmark suite.

\begin{figure*}
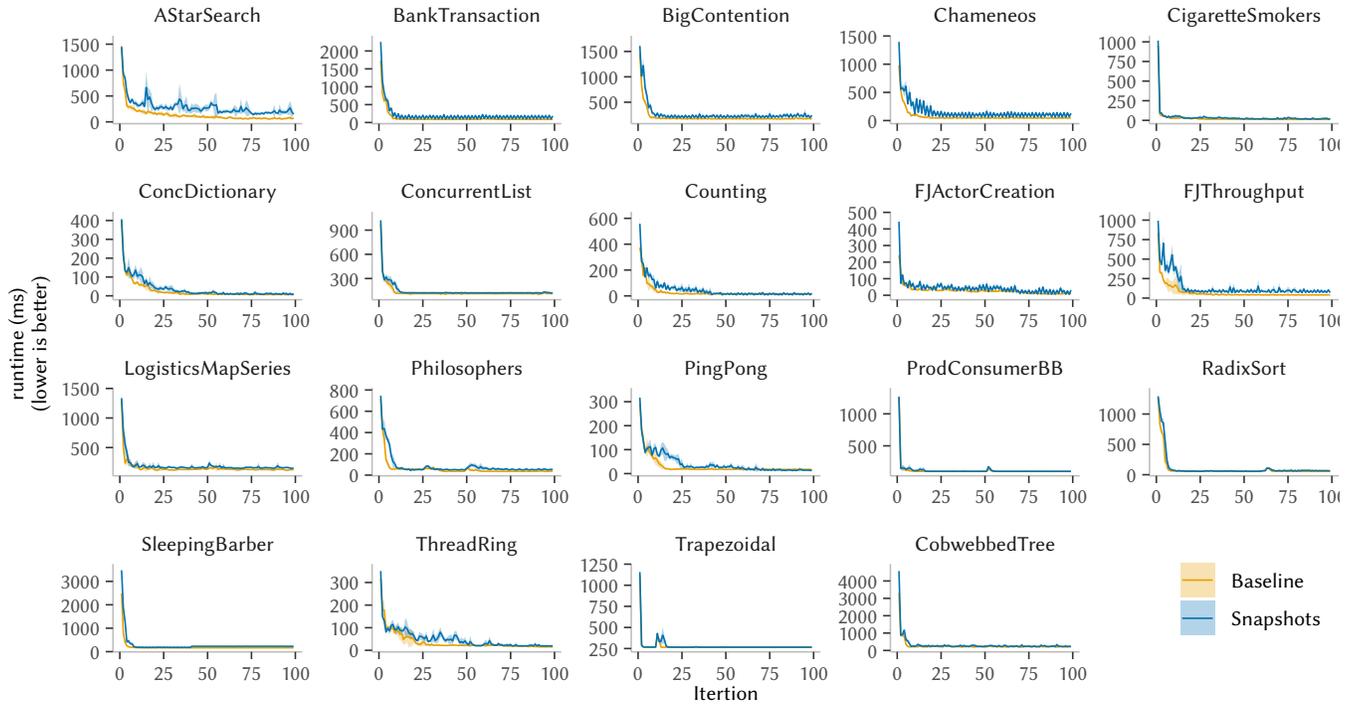

	\centering
	\SavinaWarmup
	\caption{Warmup behaviour of Savina benchmarks with and without snapshotting.}
	\label{fig:savinawarmup}
\end{figure*}

\Cref{fig:savinawarmup} shows the development of the absolute run time of the benchmarks over the first 100 iterations, which we discarded to account for warmup. The regular spikes visible in benchmarks such as \emph{Chamenos} are caused by the creation of snapshots every second iteration.

\begin{figure*}
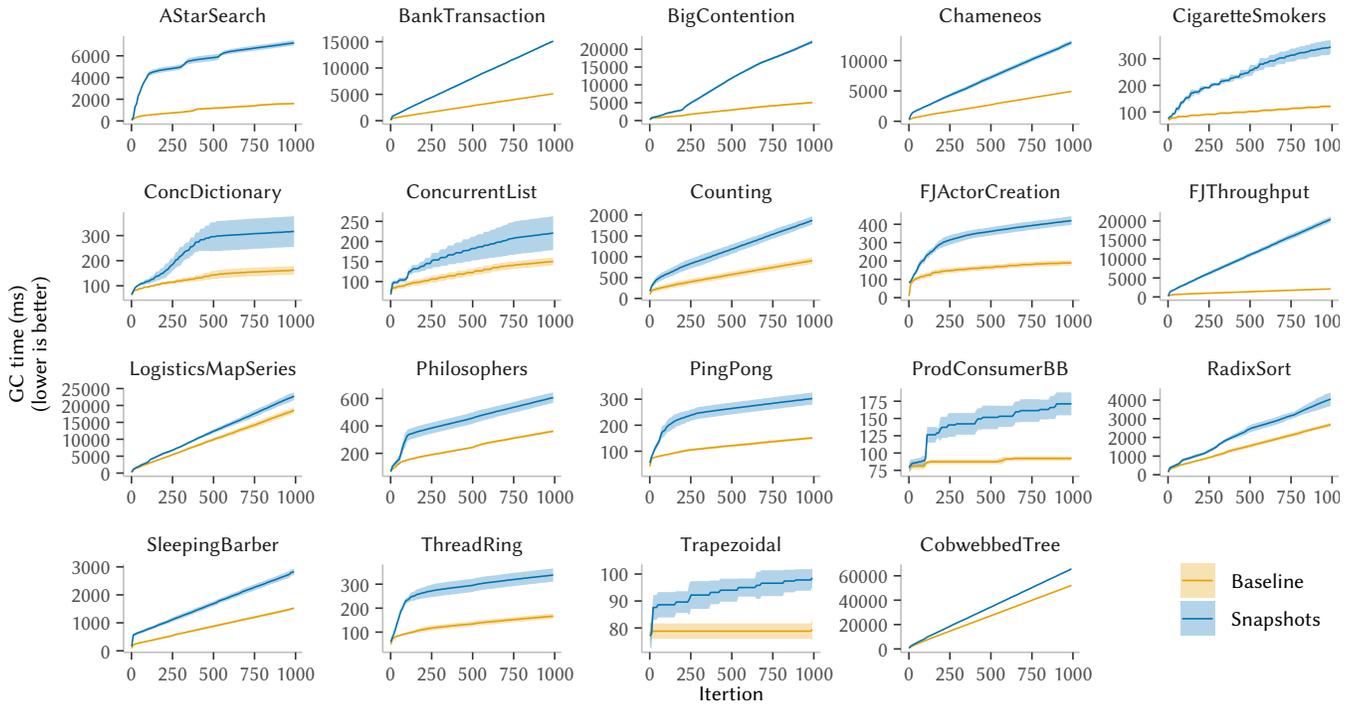

	\SavinaGCTime
	\caption{Development of total time spent on GC over 1000 iterations of Savina benchmarks, with and without snapshotting.}
	\label{fig:savinaGCTime}
\end{figure*}

\Cref{fig:savinaGCTime} shows how the GC time of the different benchmarks develops,
like the number of collected bytes and the run time overhead.
We observe that GC time varies greatly between the different benchmarks.
For some benchmarks the GC time itself contributes significantly to the measured overhead.
\emph{ForkJoinThroughput}, for example, with an average execution time of\SavinaFJTtime per iteration has a GC time overhead of\SavinaFJTGCOH per iteration.
As the average overhead of the benchmark is 2.25x, GC itself is responsible for\SavinaFJTGCOverheadShareP of the overhead.

\begin{figure*}
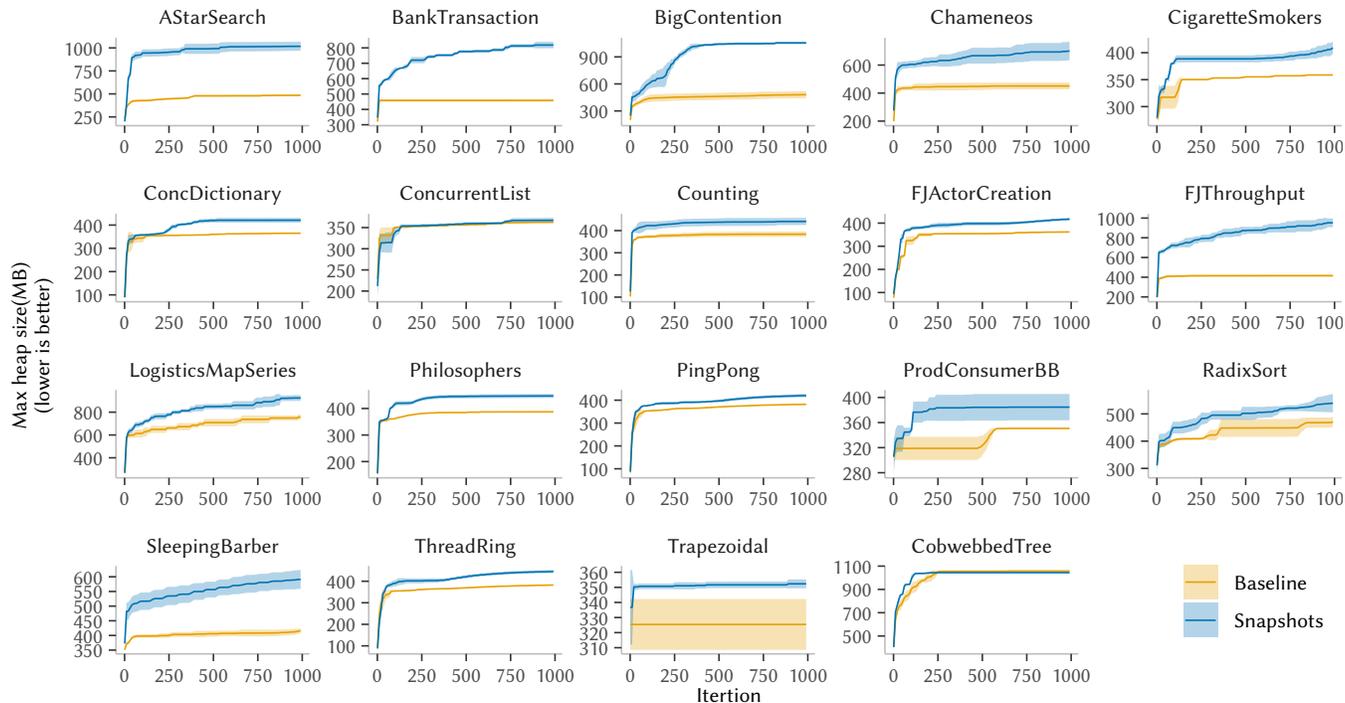

	\centering
	\SavinaHeap
	\caption{Development of the max heap size for Savina benchmarks with and without snapshotting.}
	\label{fig:savinaheap}
\end{figure*}

\Cref{fig:savinaheap} shows how the maximum heap size stabilizes over the course of 1000 iterations.
As before, there is a high variation 
between the different benchmarks. 
While benchmarks such as \emph{ConcurrentList} 
do not need additional memory when snapshotting,
benchmarks like \emph{AStarSearch} need
more than double the heap size 
and use all of the available 1GB memory.

\label{app:savina}

\end{document}